\documentclass[a4paper,fleqn,usenatbib]{mnras}

\usepackage{newtxtext,newtxmath}
\usepackage[T1]{fontenc}
\usepackage{ae,aecompl}

\usepackage[dvipsnames]{xcolor}
\usepackage{graphicx}	
\usepackage{amsmath}	
\usepackage{BibDef}

\usepackage{ragged2e}

\usepackage{booktabs}

\usepackage{multirow}

\usepackage{mathrsfs}

\usepackage{comment}

\newcommand{\msun}{\ensuremath{\, \mathrm M{\sun{}}}}
\newcommand{\rsun}{\ensuremath{\, \mathrm R{\sun{}}}}

\title[TDEs in PSB galaxies: the stellar mass function]{Tidal disruption events in post-starburst galaxies: the importance of a complete stellar mass function}

\author[E. Bortolas]{
Elisa Bortolas$^{1,2}$\thanks{E-mail: elisa.bortolas@unimib.it}
\\
$^{1}$Dipartimento di Fisica ``G. Occhialini'', Università degli Studi di Milano-Bicocca, Piazza della Scienza 3, I-20126 Milano, Italy\\
$^2$INFN, Sezione di Milano-Bicocca, Piazza della Scienza 3, I-20126 Milano, Italy\\
}

\date{Accepted 25 January 2022. Received 25 January 2022; in original form 21 December 2021}

\pubyear{2022}

\begin{document}
\label{firstpage}
\pagerange{\pageref{firstpage}--\pageref{lastpage}}
\maketitle

\begin{abstract}
A tidal disruption event (TDE) occurs when a star is destroyed by the strong tidal shear of a massive black hole (MBH). 
The accumulation of TDE observations over the last years has revealed that post-starburst galaxies are significantly overrepresented in the sample of TDE hosts. Here we address the post-starburst preference by investigating the  decline of TDE rates  in a Milky-Way like nuclear stellar cluster featuring either a monochromatic (1 $\msun$) or a complete, evolved stellar mass function.
In the former case, the decline of TDE rates with time is very mild, and generally up to a factor of a few in 10 Gyr. Conversely, if a complete mass function is considered, a strong TDE burst over the first $0.1-1$ Gyr is followed by a considerable rate drop, by at least an order of magnitude over 10 Gyr.
The decline
starts after a mass segregation time-scale, and it is more pronounced assuming a more top-heavy initial mass function and/or an initially denser nucleus. 
Our results thus suggest   that the post-starburst preference can be accounted for in realistic systems featuring a complete stellar mass function, even in moderately dense galactic nuclei. Overall, our findings support the idea that starbursting galactic  nuclei are characterized by a top-heavy initial mass function; we speculate that accounting for this can reconcile the discrepancy between observed and theoretically predicted TDE rates even in quiescent galaxies. 
\end{abstract}




\section{Introduction}\label{sec:intro}

The disruption of a star operated by a massive black hole (MBH) is known as tidal disruption event (TDE).
These  phenomena 
result  in a multiband electromagnetic emission that can exceed the MBH Eddington luminosity by orders of magnitude; in addition, TDEs feature a distinctive decline in their bolometric luminosity, that is expected to drop roughly as $t^{-5/3}$ over time-scales of months to years \citep[e.g.][]{Rees1988,Lodato2009, Lodato2011}. This makes TDEs ideal sources to unveil an otherwise quiescent population of MBHs, and their detection can provide crucial insights on the  low-mass end of the MBH mass function, since the TDE rates seem to get larger if the central MBH is less massive \citep[][]{Wang2004,Stone2016}.

The sample of observed TDEs has quickly grown to nearly a hundred in the last decades, thanks to the advent of wide-field transient surveys in the optical (as the Palomar Transient Factory, \citealt{Rau2009}; Pan-STARRS, \citealt{Chambers2016};  ASASSN survey, \citealt{Shappee2014}; the Zwicky Transient Facility, \citealt{Bellm2019}) X-ray (as ROSAT, XMM, \citealt{Jansen2001}) and gamma-ray band (e.g. \textit{Swift}, \textit{Chandra} observations). The number of detections is expected to get larger by orders of magnitude in the next years, thanks to the new observations from the presently operating eROSITA \citep[][]{Merloni2012}  and the coming  LSST/Vera Rubin  \citep{Ivezic2019} and Athena \citep{Barcons2012} observatories.

The opening of this golden era for TDEs has come with its own surprises. Together with the accumulation of the first dozens of detections, population studies of TDE hosts
have revealed that these events are significantly overrepresented in  post-starburst galaxies \citep{Arcavi2014, French2016,French2017, Law-Smith2017,Graur2018, French2020,Hammerstein2021}. The spectra of these galaxies lack strong emission lines, suggesting  low current star formation rates, but feature strong Balmer absorption lines, a signature that reveals the presence of many young stars, and it is  associated with a  starburst over the last $\approx$1 Gyr (see e.g. \citealt{French2021} for a recent review). Post-starburst galaxies are typically subdivided into the more bursty, rarer  E+A galaxies (that make up $\approx0.2\%$ of  low-redshift galaxies, and whose recent starburst created $>3\%$ of their
current stellar mass over $25-200$ Myr) and quiescent Balmer-strong galaxies (making up $\approx2\%$ of  local galaxies, and which formed $>0.1\%$ of their
current stellar mass over 25  Myr$-$1 Gyr, \citealt{French2017, French2018}).
The overrepresentation of TDEs in post-starbursts was initially  estimated to be as large as a factor $\sim 100$ for E+A  and $\sim 30$ for quiescent Balmer strong hosts \citep[][]{French2016, Graur2018}.  \citet{Law-Smith2017} showed that, once accounting for selection effects, the  E+A  post-starburst preference should be reduced by a factor $\approx 4$, bringing the overall post-starburst preference to nearly $15-30$, as later confirmed by \citet{Hammerstein2021}.

In order to investigate the origin of the post-starburst preference, it is important to properly model the dynamical processes that trigger the generation of TDEs. The most robustly understood and ubiquitous TDE production mechanism
is stellar two-body relaxation, according to which the granularity of the stellar system, and more specifically the two body interactions between pairs of stars, slowly deflect stellar orbits, some of which may eventually reach the MBH vicinity and get destroyed. \citet[][]{Stone2018} explored a set of possible dynamical drivers for the post-starburst overrepresentation, and  find that the most compelling explanation is that the host may be initially exceptionally cuspy near its centre (with density profile as steep as $\rho\propto^{-\gamma}$, $\gamma\gtrsim 2.5$, see red lines in Fig.~\ref{fig:compare_ic}), so that its initially extremely short relaxation time -- or equivalently, extremely efficient two body relaxation -- can guarantee a substantial drop of TDEs in time. However, this picture requires stellar densities to reach impressively large values about the MBH.   
Even if post-starburst galaxies  appear to harbour steep density profiles (see e.g. \citealt{Stone2016obs}; \citealt{French2020} and references therein), currently available observational campaigns can hardly resolve the MBH influence radius, and more resolved observations are required to probe the presence of the extreme cusps suggested by \citet[][]{Stone2018}. Furthermore, 
it is unclear whether star formation can proceed very efficiently so close to the MBH; molecular clouds should in fact get disrupted by its extreme  gravitational field  prior to forming stars \citep[][]{Sanders1998}. 
Alternative theories to explain the post-starburst overrepresentation are the fact that a large number of  stars can be initially set on radial orbits at the starburst stage \citep{Stone2018}; otherwise, the formation of eccentric nuclear discs about the MBH -- resulting from the gas infall accompanying the starburst --  can initially ferry a large amount of stars in the  vicinity of the MBH (\citealt{Madigan2018}; this theory   requires a large fraction of poststarburs galaxies to host eccentric discs; those could develop in galaxy mergers that then trigger the starburst, as suggested by \citealt{Hopkins2010}).

Most theoretical studies estimating dynamically induced TDE rates only account for idealized, monochromatic stellar systems composed by $1\msun$ stars. However, real stellar systems are composed by an extended stellar population that evolves with time, and accounting for this aspect could sensibly impact the dynamical modelling of TDE rates. In fact,  more massive (but generally rarer) stellar objects are expected to segregate towards the centre of the galactic nucleus, leaving lighter objects farther from the MBH, a process known as \textit{mass segregation} \citep[][]{Bahcall1977,Alexander2009, Preto2010, Alexander2017}. Mass segregation generally occurs over a fraction of the system two-body relaxation time \citep[][]{Merritt2013}, and its characteristic time-scale  can be much shorter than a Hubble time in galaxies harbouring nuclear star clusters. This process may  be very relevant for the nuclei of post-starburst galaxies, as those appear to be concentrated, and their typical galaxy and MBH masses suggest they likely host nuclear stellar clusters \citep[][]{Neumayer2020, Hoyer2021}, implying they may have relatively short mass-segregation time-scales. Furthermore, recent studies
\citep[e.g.][]{Zhang2018, Toyouchi2021} found that the stellar initial mass function (IMF) characterizing starbursting systems is generally  top-heavy (note that the Mikly Way nucleus, which is by far the best known galactic nucleus, is also characterized by a top-heavy IMF, see e.g. \citealt{Bartko2010,Lu2013}), meaning that more massive stars are formed more efficiently in comparison with the commonly adopted IMFs \citep[e.g.][]{Kroupa2001}. A top-heavy IMF can enhance the effects of mass segregation and relaxation, that ultimately set the TDE rates, and accounting for this aspect is crucial to properly track the time evolution of TDE rates.

Motivated by this, in this study we evaluate the importance of accounting for a complete, evolved initial stellar mass function in setting the time evolution of TDE rates; we also explore how varying the IMF and galaxy density profile affects the evolution of TDE rates. Those interested in a cursory reading can find a summary of the methodology at the beginning of Sec.~\ref{sec:methods} and a summary of the results in Sec.~\ref{sec:discussion}. The paper
is structured as follows: in Sec.~\ref{sec:methods} we detail the methodology adopted in this study; in Sec.~\ref{sec:results} detail our results; in Sec.~\ref{sec:discussion} we summarize and discuss our findings and present the final remarks.


\section{Methods}\label{sec:methods}

In this section we detail our implementation for addressing TDE rates. Below we briefly summarize the key aspects of our methodology, which are described in detail in the following subsections.

\begin{itemize}
    \item We adopt a Milky-Way like MBH and galaxy profile. We perform a series of runs accounting for different components: (i) the stellar bulge only; (ii) the stellar bulge and the observed nuclear star cluster (forming instantaneously at $t=0$), exploring different values for the compactness of the latter (iii) the stellar bulge and the nuclear cluster, 30 per cent of which is  formed during the simulation over a time-scale of 200 Myr.
    \item We account for a complete mass function for the system, assuming either a Kroupa IMF, or a top-heavy IMF with different power-law slopes ($\propto m^{-\alpha}$, $\alpha=\{1.5, 1.7, 1.9\}$) for masses above 0.5 $\msun$.
    \item The IMF is converted into a mass function at a given time, and large mass main sequence stars are mapped into white dwarfs (WDs), neutron stars (NSs), stellar black holes (BHs) depending on their initial mass and on their age. Stars are associated a stellar radius that sets their value for the tidal disruption radius $r_t$.
    \item Only main sequence stars can undergo TDEs in our implementation (compact objects can directly enter the MBH horizon and increase its mass, but we do not account for these events in the reported rates).
    \item Even if we account for an extended mass function, this cannot be evolved at run time in the simulation. For this, we implement a series of post-processing strategies that correct for the fact that stars age with time; note that these corrections change the event rates typically by a few per cent, and up to 25 per cent in the most extreme configurations.
    \item The system is evolved in time for 14 Gyr using the one-dimensional Fokker-Plank code Phaseflow \citep[][]{Vasiliev2017}.
\end{itemize}

\subsection{Definition of tidal disruption event}

The tidal disruption radius, i.e. the radius below which a star gets destroyed by the MBH gravitational pull, is typically written as 
\begin{equation}\label{eq:rt}
r_t = \eta \left(\frac{M_\bullet}{m}\right)^{1/3} R_\star
\end{equation}
where $m$ and $R_\star$  are the stellar mass and radius, $M_\bullet$ is the MBH mass and $\eta$ is a form factor of the order of unity that was first introduced to account for the internal structure of the star. 
In this work, we set $\eta$ to be the maximum value that guarantees a full disruption of the star. Specifically, we adopt the functional form  obtained by \citet{Ryu2020eta} 
\begin{equation}\label{eq:eta_ryu}
\begin{split}
    \eta(m, M_\bullet) = \left(0.80+0.26\sqrt{\frac{M_\bullet}{10^6\msun}}\right)\times \\
    \left(\frac{1.47 + \exp{\left(\frac{m/\msun -0.669}{0.137}\right)}}{1+2.34\exp{\left(\frac{m/\msun -0.669}{0.137}\right)}}\right);
\end{split}
\end{equation}
this quantity is of order unity, as expected (see their fig.~2). Although this is our default choice, we also test a few cases in which we fix $\eta=1$. The different choice of $\eta$ virtually  does not affect the TDE rates and their time evolution; however, we find that it can impact the relative probability of disrupting stars with different masses, as detailed in Sec.~\ref{sec:results}.

\subsection{Background galaxy density profile and star formation treatment}\label{sec:galaxy}

Our model for the background galaxy is a Milky Way model.
We implement it accounting for the stellar bulge and the central nuclear star cluster, both modelled as \citet{Sersic1968} profiles: following \citet{Pfister2020}, we adopt a bulge stellar mass of $9.1\times 10^9 \msun$ \citep{Licquia2015}, effective radius $1.04$ kpc, and Sersic index $ 1.3$ \citep{Davis2019}. If the  nuclear star cluster is present, we model it following  \citet{Pfister2020}, whose analysis is based on the observational data by \citet{Schodel2017}: the cluster   effective radius is set to 6 pc, its Sersic index to 2, and  its total mass is $ 4\times 10^7 \msun$. In all configurations, we add a central MBH of $4\times10^6 \msun$ \citep{Gravity2020} that increases its mass by accreting stellar objects.
We stress that the Milky Way properties and the MBH mass adopted for this study are similar to those inferred for many TDE hosts \citep[e.g.][]{French2020}.
The systems featuring a central nuclear star cluster expand their size over time \citep[e.g.][]{Merritt2013, Vasiliev2017};  it is thus likely that the  nuclear star cluster we observe today was initially more concentrated. For this reason, we also explore the evolution of analogous galaxies with the same properties as above but with the nuclear star cluster \citeauthor{Sersic1968} index set to $3$ and to $5$. 

In most runs we assume the nuclear star cluster (that  sets TDE rates) to be formed instantaneously  at $t=0$. This is a reasonable assumption as the starburst duration is much shorter than the age of the system (the same is assumed in \citealt{Stone2018}); in addition, the fraction of starbursting mass in post-starburst galaxies is compatible with the possibility of the entire nuclear cluster to be formed in the starburst. For comparison, we also run two cases for which  the nuclear star cluster is present and relaxed, but it undergoes extended star formation after 5 Gyr; the stellar cluster grows its mass by 30 per cent over 200 Myr ($6\times10^{-2}\msun$ yr$^{-1}$), after which star formation is quenched and the system is further evolved in time. Star formation is implemented by adding a source term in the Fokker-Plank treatment, it is constant in time and occurs within  $r\approx6$ pc, with the form  $\dot{\rho}\propto r^{-1/2}$ (see \citealt{Vasiliev2017} for more details).

\subsection{The IMFs}
\label{sec:imf}
We sample stars in the range [0.07,150] $\msun$.
Our fiducial IMF choice is the  \citet{Kroupa2001}; together with it, we  explore the effect of top-heavy IMFs. All our IMFs $\chi(m)$ are broken power laws with the form
\[
\chi(m) \propto
\begin{cases}
    m^{-1.3}  & \text{if $m<0.5\msun$},\\
    m^{-\alpha}  & \text{if $m\geq0.5\msun$};
\end{cases}
\] 
with  $\alpha=2.3$ for the \citeauthor{Kroupa2001}, and $\alpha=\{1.5, 1.7, 1.9\}$ for the top-heavy IMFs. 

\subsection{Stellar evolution and stellar parameters}\label{sec:stellar_evolution}

We now describe a series of prescriptions  adopted to map the initial mass of stars into their mass and radius at any given time. Note that the age of the stellar population needs to be set at the beginning of the integration, but it remains constant over the course of the simulation, meaning that stellar masses and radii are not evolved at run time. However, we adopt a series of post-processing routines to correct for this simplification, as described in Sec.~\ref{sec:stellar_evolution_corrections}. 

\subsubsection{Main sequence stars}

We assume main sequence stars to evolve negligibly in mass and radius over time, and we assign them a radius following \citet[][and in particular their appendix A1]{D'Orazio2019}  who approximate the prescriptions by \citet{Tout1996}:
\begin{equation}
R_\star = \left(\frac{m}{\msun}\right)^\xi \rsun
\end{equation}
with $\xi = 0.85$ if $m<1\msun$ and $\xi = 0.6$ if $m\geq 1\msun$.
Main sequence stars undergo a TDE if they approach the MBH at a distance smaller than $r_t$; if so, 30 per cent of their mass is added to the central MBH, while the remaining is assumed to be lost in the interstellar medium and as radiation (implying that the total mass  is not conserved in the run).

Stars exit the main sequence to directly become compact remnants (WDs, NSs, BHs)
if their main sequence mass is larger than $m_{\rm break}$, which here is assumed to depend on the age of the stellar population $t_\star$ via
\begin{equation}
\label{eq:time_mass}
    \frac{m_{\rm break}}{\msun} = \left(\frac{t_\star}{\rm 10\ Gyr}\right)^{-1/2.5}.
\end{equation}

\subsubsection{Compact stellar remnants}

Stars whose initial mass is below 8.70 $\msun$ are expected to end their lives as WDs.  The initial-final mass relation for WDs 
is taken from the analytical prescription described in \citet{Cummings2018}, in particular their fit to the PARSEC \citep{Bressan2012, Tang2014, Chen2014nov} evolutionary tracks:
\begin{equation}
m_{\rm WD}= 
\begin{cases}
    0.0873  m_{\rm MS}+ 0.476 \msun& \text{if } 0.87 \msun < m_{\rm MS} \leq 2.80 \msun\\
    0.181  m_{\rm MS}+0.210\msun & \text{if } 2.80 \msun < m_{\rm MS} \leq 3.65 \msun\\
    0.0835  m_{\rm MS}+0.565\msun & \text{if } 3.65 \msun < m_{\rm MS} < 8.20 \msun
    \label{eq:WDmass}
\end{cases}
\end{equation}
here $m_{\rm WD}$ is the mass of the WD and $m_{\rm MS}$ is its progenitor mass in the main sequence. In the current work, the mass of the  MBH is always too large to allow for tidally disrupting a WD, which would instead enter the MBH horizon prior to its disruption.


Stars whose mass is initially above $8.7\msun$ are mapped into the mass of the associated  compact remnant (NS, BH). 
We map their initial mass into their final mass adopting the PARSEC evolutionary tracks at Solar metallicity \citep{Bressan2012, Tang2014}; the mass of the compact remnant is obtained according to \citet{Spera2015}, assuming the delayed supernova model; the initial to final mass of such massive stars can be seen in fig.~1 of \citet{Bortolas2017}. 

WDs, NSs and BHs  are lost into the MBH (and all their mass is added to the MBH) if they get closer than $r_\bullet = 4GM_\bullet/c^2$  
(here $G$ is the gravitational constant, and $c$ is the speed of light in vacuum). These events may give rise to detectable gravitational wave signals (the so-called EMRIs, see e.g. \citealt{Zwick2020, Zwick2021, Vazquez-Aceves2021}) but they are not TDEs.\footnote{The supernova natal kick experienced by NSs and BHs can also funnel them within $r_\bullet$ \citep[][]{Bortolas2019}, but we did not account for this possibility in the present study.}

\subsubsection{Mass families in the Fokker-Plank implementation}

The Fokker~Plank code adopted for the evolution of the system does not allow us to include a continuous mass function. For this,  we divide the initial stellar mass range into 100 logarithmic bins,\footnote{We performed a convergence study varying the number of mass bins, and we found our choice to differ by only 0.1 per cent in TDE rates compared with a case with a sampling of 500 bins.} each characterized by its  initial  and evolved stellar mass, number of objects, stellar radius and so forth. Each mass bin is called a \textit{mass family}. 

\subsubsection{Post-processing corrections for stellar evolution}\label{sec:stellar_evolution_corrections}

Our choice for $m_{\rm break} $ is $ 2 \msun$ (i.e. the population is $\approx 1.8$ Gyr old; see the beginning of  Sec.~\ref{sec:stellar_evolution});  all mass bins characterized by an initial stellar mass exceeding this value are treated as compact objects during the simulation. We checked that our specific choice of $m_{\rm break}$ does not significantly impact the results of the present work.
We implement two post-processing corrections that account for the fact that the stellar population starts out younger and eventually gets older than the time associated with $m_{\rm break}$. 

First of all, for $t>t_\star(m_{\rm break})$, we neglect all TDEs that are associated with a mass family that already became a compact stellar remnant  at that given $t$. Secondly, for $t<t_\star(m_{\rm break})$, we estimate the TDE rate that would be associated with each $k$-th mass family with initial mass larger than $m_{\rm break}$ as
$\dot{N}_k = \dot{N}_{\rm ref} (n_k/n_{\rm ref})$; this allows to correct for the fact that in the code we cannot destroy very massive stars that quickly become compact remnants. Here $\dot{N}_k, n_k$ respectively represent the TDE rate and number of objects in the system associated with the $k$-th mass family, while $\dot{N}_{\rm ref}, n_{\rm ref}$ represent the same quantities associated with a reference mass family that can be used as a proxy for the initially heavier $k$-th family. Here we choose the reference mass family to be the largest stellar mass still in the main sequence (2$\msun$). Both the implemented corrections are small, and together they never change the event rates by more than $\approx25$ per cent (the maximum correction is associated with the $\alpha=1.5$ IMF).

\subsubsection{Correction for the total mass of the system}\label{sec:correctiontotalmass}

Even \textit{neglecting TDEs } and focusing only on the stellar population, it is important to caution that  its total mass declines in time, as main sequence stars are turned into compact objects of lower mass; for a given a system, the conserved quantity is the total number of objects. In order to consistently compute the TDE rates within a given system and normalize its total mass consistently for all the chosen IMFs, here we opt for  defining the total mass of the system (Sec.~\ref{sec:galaxy})
when the stellar population has a specific age. In what follows, we always define the total mass at $t_\star$=10 Gyr, or equivalently at $m_{\rm break, ref} =1\msun$, meaning that the total mass is set for a 10 Gyr old population, and the real total mass  evolved in the simulation is (13, 7, 10, 13) per cent larger  for a (Kroupa, $\alpha=1.5, 1.7, 1.9$) IMF. We do not apply any correction when we treat a monochromatic mass function.

\subsubsection{Further approximations in the treatment of the mass function}

In this paper, we do not account for the presence of binaries and multiplets. The larger effective mass of these systems compared to normal stars make them subject to  mass segregation, and neglecting them may result in slightly underestimating the effectiveness  of mass segregation. However, we expect this effect to be relatively small, and we believe the inclusion of multiplets can potentially enhance even further the early TDE rates.

Concerning giants, albeit their very large radii could in principle make them ideal candidates for TDEs, their low density stellar envelope makes them prone to less violent and thus more elusive events \citep[e.g.][]{Rossi2020}, as the so-called ``spoon-feeding'' \citep{MacLeod2013}. In addition, the giant stages are typically short compared to the other stages considered here. For this, we do not account for giants in the present implementation, and we assume stars whose mass is above $m_{\rm break}$ to directly turn into compact remnants.

\subsection{Fokker-Plank integration}\label{sec:phaseflow}

The evolution of the system is performed by means of the Phaseflow Fokker-Plank integrator \citep[][]{Vasiliev2017}, which is part of the AGAMA toolkit \citep[][]{Vasiliev2019}. The code allows us to evolve in time an isotropic, spherically symmetric distribution of stars  by solving the coupled system of Poisson and orbit-averaged Fokker–Planck one-dimensional equations for the gravitational
potential, the density, and the distribution function. The evolution of the system is driven by two-body relaxation and mass segregation. The system can account for the presence of a sink term representing the loss of objects into the MBH (whose mass is allowed to grow, and $r_t$, $r_\bullet$ are consistently adjusted accordingly) and possibly a source term that mimics star formation. The code also has the capability of evolving multiple stellar \textit{families} with their own properties (described above).
The code has been successfully tested in the context of e.g. nuclear star clusters evolution \citep{Generozov2018, Emami2020}, TDEs in massive high redshift clumps \citep{Pestoni2021} and primordial BHs \citep{Zhu2018, Stegmann2020}.

In this work we  set the simulation parameters as follows: the phase-volume grid is chosen to have 400 logarithmically spaced points; we select an integration accuracy of $10^{-3}$ and the quadratic discretization method. The Coulomb logarithm for the evolution of the system is set to  $\ln\Lambda = \ln\left(0.4M_\bullet/\langle m\rangle \right)$  \citep{Spitzer1971}, with $\langle m\rangle$ defined in Eq.~\ref{eq:mavg}.

\subsection{Static estimates of TDE enhancements}

The  TDE rate variation associated with adopting a complete, instead of a mono- or bi- chromatic stellar mass function, has been first explored by \citet{Magorrian1999}, and subsequently addressed by \citet{Kochanek2016, Stone2016, Pfister2022};\footnote{\citet[][]{Foote2020} instead focus on the effect of mass segregation and TDE rate enhancement in eccentric nuclear discs with a bi-chromatic mass function, but they do not explore the more general spherical and isotropic scenario and the effect of a complete mass function.} in particular, \citet{Stone2016} presented a simple back of the envelope estimate that we repropose and generalize here, based on the stellar  evolution framework described above. 
The rate change in TDEs when considering a complete mass function, rather than a  monochromatic mass function of 1 $\msun$ stars, can be estimated by accounting for a series of   effects.

First of all, for a \textit{fixed} total (stellar) mass of the system, a different mass function implies that the total number of stars in the system would be different; in particular, the number of objects in a galaxy with given total mass scales as  the inverse of the average mass
\begin{equation}\label{eq:mavg}
\langle m\rangle = \int_{m_{\rm min}}^{m_{\rm max}} \chi(m)\, m_f(m) dm,
\end{equation}
where $m_f(m)$ is the function mapping the initial in the current stellar mass.
A second effect is related to the angular momentum diffusion coefficients; those  eventually set the TDE rate in the  empty loss cone regime (see Sec.~\ref{sec:mass_preference} for its definition), and they  are  $\propto \ln\Lambda \times \langle m^2\rangle /\langle m\rangle$, where $\langle m^2\rangle$  is the second moment of the mass function
\begin{equation}
    \langle m^2\rangle =\int_{m_{\rm min}}^{m_{\rm max}} \chi(m)\, m_f^2(m) dm
\end{equation}
and $\ln\Lambda$ is the Coulomb logarithm defined in Sec.~\ref{sec:phaseflow}.
Finally, not all objects in the system can give rise to TDEs, as their disruption radius can be smaller than the MBH horizon (e.g.  WDs, NSs, BHs).  
The \textit{static} TDE rate enhancement that characterizes a complete mass function, compared with a monochromatic stellar population of $1\msun$ stars can thus be estimated as
\begin{equation}\label{eq:ratefrac}
    \frac{\dot{N}_{\rm MF}}{\dot{N}_{\rm mc}}\approx \frac{\langle m^2\rangle}{\langle m\rangle^2} \frac{\ln({0.4 M_\bullet/\langle m \rangle})}{\ln({0.4 M_\bullet/\msun})}\int_{m|r_t>r_\bullet} \chi(m) dm;
\end{equation}
${\dot{N}_{\rm MF}}$ and ${\dot{N}_{\rm mc}}$ respectively represent the TDE rates associated with a population with extended and monochromatic mass function. The  integral  is meant to reduce the number of events excluding stellar objects that are too compact to undergo a TDE; in practice, here the integral should be carried out between $m_{\rm min}$ and $m_{\rm break}$.

In order to properly normalize the total mass of the system at any given time, we estimate the ratio in Eq.~\ref{eq:ratefrac} accounting for the correction described in Sec.~\ref{sec:correctiontotalmass}. 
This practically means that  $\langle m\rangle$ appearing at the denominator (which in fact tunes the number of objects in the system) should be evaluated considering a population whose $m_f(m)$ is evaluated for a population of $t_\star$=10 Gyr, while all the other quantities are computed at the chosen time of the stellar population.
The rate enhancements as computed from Eq.~\ref{eq:ratefrac} are equal to respectively $\{7.3, 8.7, 7.9, 4.3\}$ for IMFs with $\alpha=1.5,1.7,1.9$ and the Kroupa IMF; those values are evaluated for $t_\star = $1.8 Gyr, corresponding to $m_{\rm break}=2\msun$.\footnote{The introduced static estimate is  valid in the empty loss cone regime, and it neglects the fact that stars have a different $r_t$ depending on their mass. \citet[][]{Kochanek2016} introduce a more elaborate static correction for the presence of a complete mass function, based on the previous works by \citet[][]{Wang2004} and \citet{Magorrian1999}; in particular, according to their eq.~3, the rate enhancement compared with a monochromatic mass function would be
\begin{equation}\label{eq:ratefrac_kochanek}
    \frac{\dot{N}_{\rm MF}}{\dot{N}_{\rm mc}}\approx \frac{{\langle m^2\rangle}^{3/8}}{\langle m\rangle^2}
    \int_{m|r_t>r_\bullet} \left(\frac{R_\star(m)}{R\odot}\right)^{1/4}  \eta(m)^{1/6} m^{-1/12} \chi(m) dm.
\end{equation}
The  rate enhancements computed via the  equation above within our stellar evolution framework are very modest, in the range $1-1.6$, and they are in reasonable agreement with what found by \citet[][]{Kochanek2016}; however, they fail to reproduce the rate enhancements obtained with Phaseflow even at the very beginning of the integration.}  
It is important to note that the estimate neglects the time evolution of the system, the fact that stars have a different $r_t$ depending on their mass and the fact that  mass segregation affects  the evolution of the system. The same  values computed from the PhaseFlow  integrator  (whose  rates are presented in Sec.~\ref{sec:results})  at $t=0$ are in the range ${\dot{N}_{\rm MF}}/{\dot{N}_{\rm mc}}=3-6$ if the nuclear cluster Sersic index is set to 2, and in the range ${\dot{N}_{\rm MF}}/{\dot{N}_{\rm mc}}=4-7$  if it is set to 5. The presented rough estimate is thus not too bad (even if necessarily approximate) at the beginning of the evolution, but it becomes completely unreliable as time passes, as shown below.

\section{Results} \label{sec:results}

\subsection{TDE rates from the stellar bulge}

First of all, we address the TDE rates for a galaxy featuring the stellar bulge only, without a nuclear stellar cluster. We evolve the system, modelled with a Kroupa IMF, for 14 Gyr. The density profile of the system at the end of the integration is virtually indistinguishable from the initial conditions, suggesting relaxation is too weak to impact the evolution of the system. Accordingly, the TDE rate  remains very low,  nearly $\approx 2.3\times 10^{-7} \ {\rm yr}^{-1}$ at all times. 
This implies the bulge contribution to the TDE rates is negligible compared to that of the nuclear star cluster, reported in the next section (in agreement with \citealt{Pfister2020}). This also means that TDEs can be assumed to come from stars in the nuclear star cluster only to a very good approximation, while the bulge virtually acts only as a background potential.


\subsection{TDE rates accounting for a nuclear star cluster}

\begin{figure}
\centering
\includegraphics[ width=0.44\textwidth]{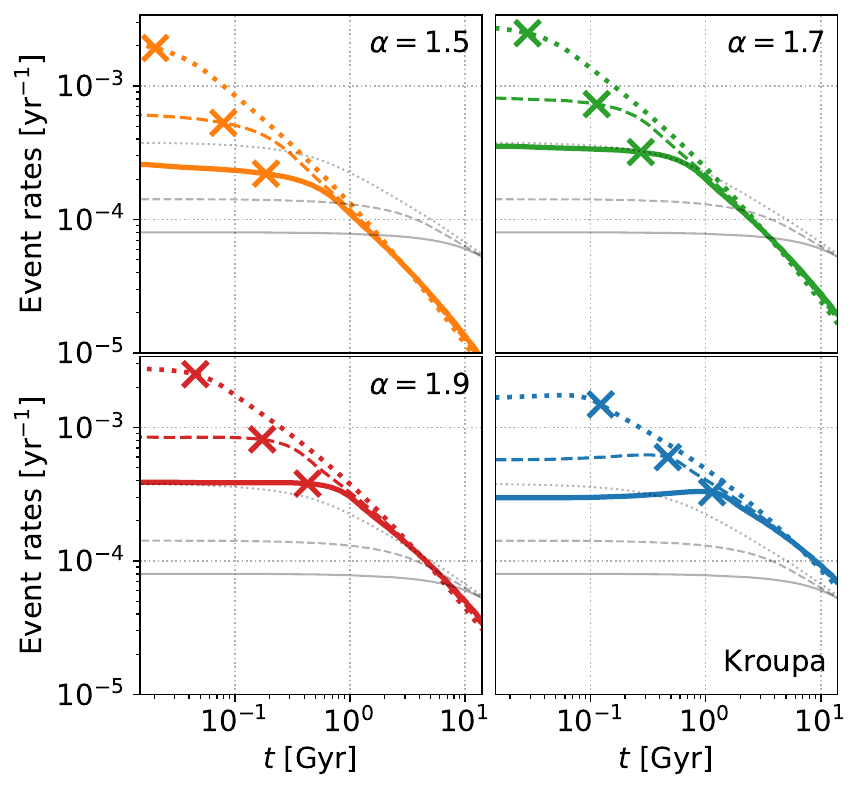}
    \caption{The panels show the time evolution of TDE rates for different choices of the background system and IMF.
    The top left, top right, bottom left, bottom right panels respectively show the evolution of the TDE rates  associated with a top heavy IMF with  $\alpha=1.5, 1.7,1.9$ and a Kroupa IMF. In all panels, the thin grey lines show the TDE rates associated with a monochromatic IMF (1$\msun$ stars). The solid, dashed, dotted lines in each plot refer to nuclear star clusters whose Sersic index is equal respectively to $2, 3, 5$. 
    \textit{The TDE rates significantly decline at late times for all the chosen IMFs, except for the monochromatic one.}
    The `\textit{x}' on the top of each coloured line marks the associated value of the mass segregation time-scale, computed via Eq.~\ref{eq:tms}. The event rates start dropping significantly in all runs after this time-scale.}  
    \label{fig:time_segregation}
\end{figure}

\begin{table}
\centering
\begin{tabular}{cccccc}
\hline
Sersic  &  &  Top-heavy& &\\
index  & $\alpha=1.5$  & $\alpha=1.7$ & $\alpha=1.9$ & Kroupa & Monochr.  \\
\hline
 2 & 29  (14.5) & 19  (11.3) & 11 (8.1)   &  5  (3.7)  & 2 (1.3)\\ 
 3 & 77  (19.7) & 47  (16.3) & 26 (12.8)  &  9  (6.9)  & 3 (2.2)\\
 5 & 312 (24.9) & 178 (21) & 91 (16.6)  &  27 (9.4)  & 7 (4.5) \\
\hline
\end{tabular}
\caption{The table displays the ratio between the initially large event rates and the lower, late event rates for the runs shown in Fig.~\ref{fig:time_segregation}; each run is  defined by its IMF (columns) and nuclear cluster Sersic index (rows). In particular, numbers out of the parenthesis show the ratio between the maximum and minimum value of TDE rates for each given run. The numbers in parenthesis, instead,  show the ratio between the early TDE rate computed between $250-750$  Myr, and the late event rate between $10-12$ Gyr. The inclusion of a complete, evolved IMF induces a significant decline of the TDE rate with time; the same is not observed adopting a monochromatic mass function.}
\label{tab:time_enhancements_nsc}
\end{table}

Fig.~\ref{fig:time_segregation} shows the time evolution of the TDE rates for a set of models that account for both the bulge and the nuclear star cluster. The results are shown for different IMFs and for nuclear cluster \citeauthor{Sersic1968} index equal to 2, 3, 5; in all panels, we  also show the TDE rates associated with a monochromatic mass function. Remarkably, in all cases, the TDE rates significantly drop over time if one adopts an extended mass function, while the drop is very mild in the monochromatic case. Tab.~\ref{tab:time_enhancements_nsc} reports the peak to valley ratio for the event rate in each run; it also shows the ratio between the average event rate at $t = 250-750$ Myr and the same value at $t = 10-12$ Gyr. Except for the monochromatic case,  the early event rates are significantly larger than the late time ones ($\sim\times10-100$, except for the less concentrated Kroupa case). For this specific choice of galaxy and MBH mass, the early to late TDE rates are thus compatible with the post-starburst preference (which is by nearly a factor $15-30$)  especially if considering top-heavy IMFs and/or initially more concentrated clusters.\footnote{During the 14 Gyr of evolution, the MBH reaches a mass of $4.4-5.1\times10^6\msun$ depending on the chosen initial conditions, with a more efficient growth in case of higher initial Sersic index and lower value of $\alpha$.}

We stress that the different TDE rates for varying IMFs here are obtained for systems with the same mass (at $t=10$ Gyr). In order to properly compare these results to observations, one should in principle normalize different systems depending on their luminosity in a given band, that necessarily depends on the chosen IMF and age of the stellar population. This is beyond the scope of the present work, but accounting for this would necessarily partially affect the TDE rates normalization and time evolution.

\subsection{Mass segregation as the driver of the evolution}

In order to best interpret the results described so far, it is important to recall that an important prediction of the relaxation theory is that stars orbiting an MBH redistribute in a universal mass profile over $0.1-0.25$ relaxation times:
they develop a Bahcall-Wolf density cusp $\rho(r)\propto r^{-\gamma}$ within the MBH influence
radius, with $\gamma$ = 7/4 in the monochromatic case \citep[][]{Bahcall1976}.
If the system is constituted by a composite stellar population, the heavier objects (BHs, NSs) tend to settle on a moderately steeper cusp with $\gamma = 1.75-2$, while low mass stars develop a weaker cusp with $\gamma\approx 1.3-1.5$ \citep{Bahcall1977,Preto2010, Amaro-Seoane2011, Merritt2013}. This is nothing else than a manifestation of mass segregation, i.e. the tendency of  massive objects  move closer to the centre of the system, leaving lighter objects outwards.
Following \citet[][]{Merritt2013}, the mass segregation time-scale --  i.e. the time over which the segregation of heavier stars can be considered to be completed -- can be estimated as
\begin{equation}\label{eq:tms}
    t_{\rm MS}(r) \approx \frac{0.0814\, \sigma^3(r)}{G^2 \Tilde{m} \rho(r) \ln \Lambda}
\end{equation}
where  $\sigma(r)$ and $\rho(r)$ are the one-dimensional velocity dispersion and stellar density of the system at that given radius, while  $\Tilde{m} =\langle m^2\rangle /\langle m\rangle$. We compute this time-scale in the initial conditions of the runs by extracting the value of $\rho(r),\, \sigma(r)$ in the assumption of spherical symmetry and isotropy from the AGAMA toolkit \citep[][]{Vasiliev2019}. $t_{\rm MS}(r)$  has a minimum close to $\sim 1$ pc, which is reasonable to use as the reference $t_{\rm MS}$. This time-scale  is marked with an $x$ for each model shown in Fig.~\ref{fig:time_segregation}. From the plots, it is apparent that the mass segregation time-scale sets the time span after which the event rates start declining.

\begin{figure}
\centering
\includegraphics[ width=0.45\textwidth]{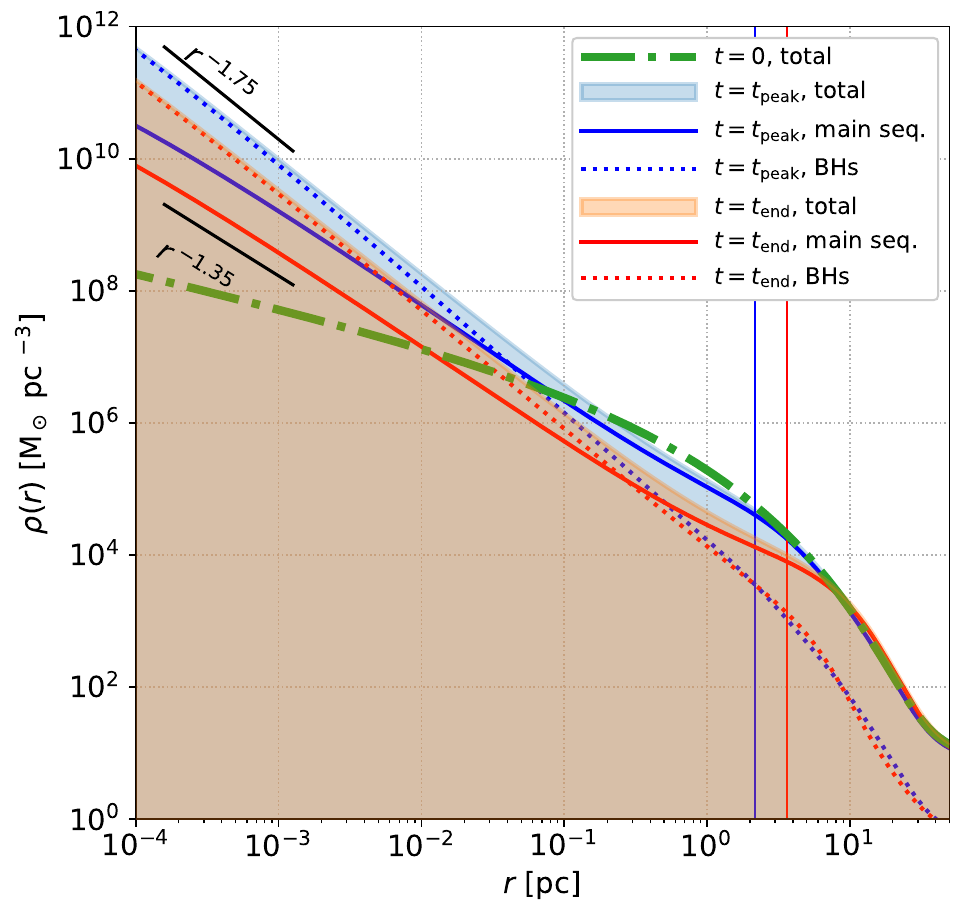}
    \caption{Density profile of the system initialized with a Kroupa IMF and with nuclear cluster Sersic index set to 2 (the density evolution in all the other runs is analogue to this). The green dash-dotted line shows the total initial density profile. The blue and red curves respectively refer to the profiles at the moment the density cusp is fully developed, and at the end of the run (14 Gyr). The solid line refers to the density profile of main sequence stars, the dotted line refers to the profile of compact stellar remnants with mass above 2$\msun$ (thus mainly BHs), and the upper edge of the shaded areas show the total density profile. The vertical, thin solid lines mark the radii at which the TDE flux is maximum. The black segments show as a reference the power laws $\rho\propto r^{-1.75}$ (top) and  $\rho\propto r^{-1.35}$ (bottom). The density profile for main sequence stars and BHs respectively follow those  power-law trends in the inner region to a good approximation.}
    \label{fig:density_evolution}
\end{figure}

Before $t_{\rm MS}$, mass segregation is not yet completed, implying that the cusp is growing about the MBH both for heavier and lighter stellar objects; as a reference, Fig.~\ref{fig:density_evolution} shows the density profile (for BHs, main sequence stars, and total) at the beginning, at the moment the cusp is fully grown ($t\approx 1.8$ Gyr) and at the end of the evolution for the run with a Kroupa IMF and Sersic index set to 2. After the cusp is fully grown, the system further relaxes, expanding (thus lowering the normalization of the density) in time as a result of the central MBH acting as an heat source \citep[][]{Vasiliev2017}.  The system expansion and density drop is more efficient in systems featuring a top-heavy IMF, as relaxation is more efficient.  Fig.~\ref{fig:density_evolution} also marks the radius from which most TDEs are generated. Once segregation is completed, this radius tends to move outwards, where the density is lower. The event rates thus drop after a mass segregation time-scale as the radius from which the TDE flux is maximum  gets  larger, and the density at all radii gets lower. The presented picture remains nearly the same for all runs.

In all explored cases, the late time event rates are very similar if one assumes the same IMF, regardless of the initial cluster profile. This is  due to the fact that the system density profile relaxes to a very similar  density profile, regardless of the initial condition, and whose normalization slowly drops as time passes; if the system is initially more concentrated, its initial TDE rates are larger, but the system expands more efficiently, so that the two effects compensate each other and the final rates are similar.

\subsection{Preference for the disruption of more or less massive stars}\label{sec:mass_preference}

\begin{figure}
\centering
\includegraphics[ width=0.38\textwidth]{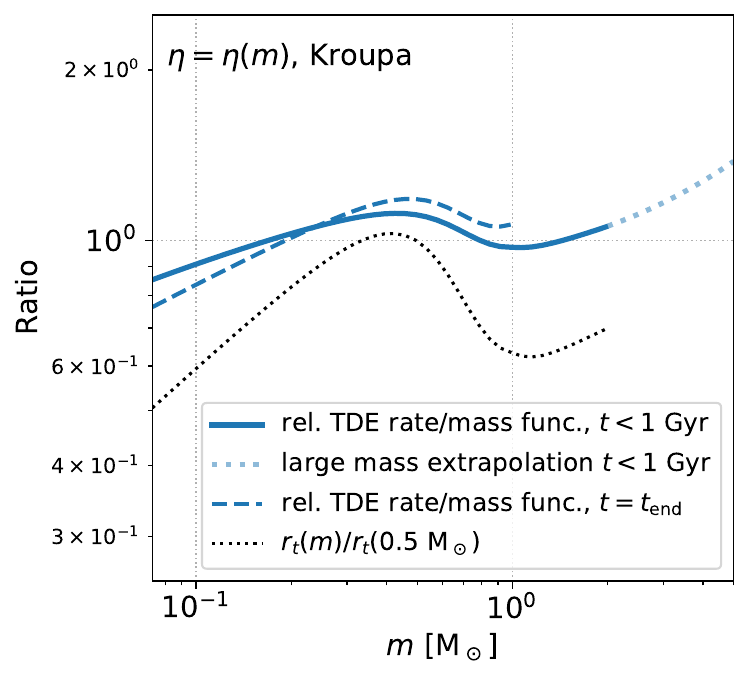}
\includegraphics[ width=0.38\textwidth]{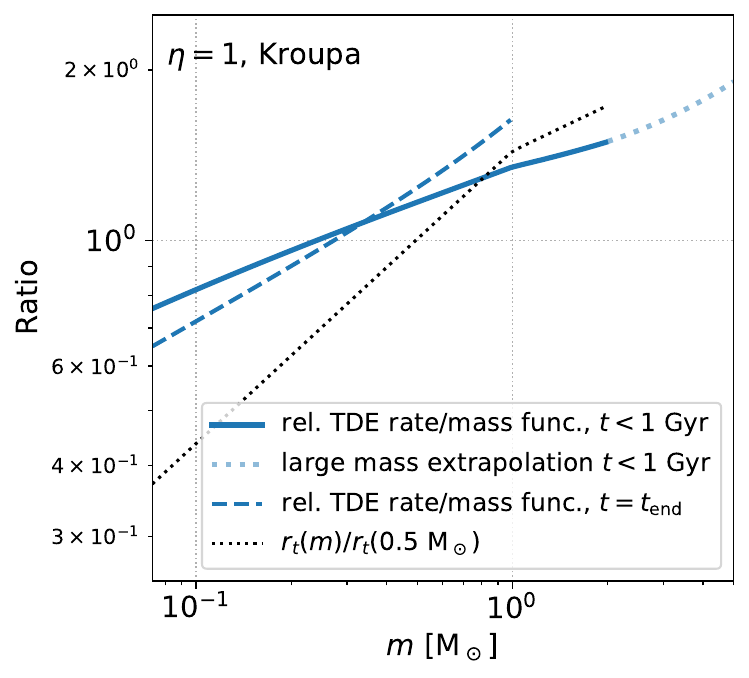}
    \caption{The plots shows the probability of disrupting a star of mass $m$, normalized to the number of stars with that given $m$ in the IMF. We show this value for TDEs occurring over the first Gyr (solid line) and in the final snapshot (14 Gyr, dashed line); in the former case, we show an extrapolation for large masses with a dotted line -- see footnotes~\ref{fn:extrap_eta},~\ref{fn:extrap_no_eta} for more details; in the latter, we set the event rates for stars with $m>1\msun$ to 0 as those would have become WDs. For reference, we show the tidal disruption radius $r_t$ arbitrarily normalized to its value for a star of $0.5 \msun$ with a dotted black line. The top plot refers to the run with TDE radius defined via Eq.~\ref{eq:eta_ryu}; the bottom one assumes $\eta=1$. Both plots show the runs with a Kroupa IMF and Sersic index equal to 2.}
    \label{fig:rel_rate_mass}
\end{figure}

The complete mass function evolved in our runs allows us to track which stars are more likely to be disrupted as a function of their mass. An educated guess would be that stars of a given (initial, or equivalently here, main sequence) mass are destroyed proportionally to their IMF, i.e. $\dot{N}(m)/\dot{N}\propto\chi(m)$, where $\dot{N}(m)$ is the TDE rate for main sequence stars with mass $m$, $\dot{N}$ is the total TDE rate [$\dot{N}=\int \dot{N}(m)dm$], and again $\chi$ represents the stellar mass function. Based on this, Fig.~\ref{fig:rel_rate_mass} (top panel) shows the ratio $\tfrac{\dot{N}(m)}{\dot{N}}/\chi(m)$;\footnote{Here $\chi(m)$ has been normalized to 1 between the minimum allowed mass and $m_{\rm break} = 2\msun$, in order to avoid accounting for compact objects that cannot undergo a TDE.} we show this ratio for the early ($t<1$ Gyr) and late ($t$=14 Gyr) evolution.\footnote{\label{fn:extrap_eta}We also extrapolate the ratio for the large stellar masses that would be present at early times by performing a linear fit of the shown data between $m=1.3\msun$ and $m=2\msun$, obtaining Ratio$(m)=0.85\msun+0.11m$ (this will be used in the next Subsection).}
The plotted ratio remains close to 1 for all values of $m$, and it is interesting to notice that the shape of the residual curve is very similar to the trend of the tidal disruption radius $r_t(m)$, also plotted in the Figure with a dotted black line. This suggests that the value of $r_t$ associated to each mass $m$ also has an impact on the TDE rates. 

In order to better explore this aspect,  we  also perform an additional run in which we fix  $\eta=1$ (the value that defines the TDE radius, see Eq.~\ref{eq:rt}) instead of allowing it to vary with $m$ (as in the standard runs, Eq.~\ref{eq:eta_ryu}). The case with $\eta=1$ is shown in the bottom panel of Fig.~\ref{fig:rel_rate_mass}. Again, the residual curve resembles the trend of $r_t$, confirming that the choice of the TDE radius   has an impact on the number of events to be expected for each given $m$.\footnote{\label{fn:extrap_no_eta} In this case, the large mass linear regression between 1.3 and 2 $\msun$ results in Ratio$(m)=1.2\msun+0.14m$.}  A proper definition of $r_t$  is thus crucial to properly compare observations and theoretical work.
We checked that the results presented in this Subsection  remain very similar if one varies the IMF or the Sersic index of the nuclear stellar cluster.\\

The result presented in this Subsection can be understood in terms of the loss cone theory. In short, the loss cone is the region in the phase space containing stars that at any given time have their periapsis smaller than their tidal disruption radius, thus they are good candidates for undergoing TDEs. The space about the MBH can be divided into two  regions: closer to the MBH, stars  are in the so-called \textit{empty loss cone regime}, meaning that the stellar orbital periods are much smaller than the typical time-scale over which a star enters or exits the loss cone; as a consequence,  stars in the loss cone are immediately destroyed by the MBH. Farther away, stars are  in the \textit{full loss cone regime}, meaning that the orbital period of stars gets much longer than the typical time-scale over which a star can enter and exit the loss cone as a result of two body relaxation; here the loss cone is always full, and stars in there can enter and exit it many times before being disrupted (see e.g. \citealt{Merritt2013}). It is customary to define the energy- (or radius-) dependent parameter
\begin{equation}
    q= \frac{\mathcal{D} T_{\rm orb}}{(l_{\rm lc}/l_0)^2}
\end{equation}
where $\mathcal{D}$ is the diffusion coefficient in angular momentum (note that $\mathcal{D}^{-1}$ is a good approximation for the relaxation time-scale), $T_{\rm orb}$ is the orbital period (which is nearly independent of the angular momentum, and can be approximated as a sole function of the energy),  $l_0$  represents the angular momentum  of a circular orbit at that energy, and   $l_{\rm lc}\approx \sqrt{2GM_\bullet r_t}$ is the loss cone angular momentum of a star whose periapsis is equal to $r_t$. Since $\mathcal{D}^{-1}(l_{\rm lc}/l_0)^2$ is a good estimate of the time-scale needed by a star of that given energy and angular momentum equal to $l_{\rm lc}$ to have its angular momentum changed by $\approx l_{\rm lc}$ due to relaxation, $q$ is in fact a measure of relaxational loss cone refilling at the energy (or radius) at which it is computed; if $q\ll1$ ($\gg1$) for a given energy (or radius), then the loss cone is empty (full).

In the presented runs, the  value of $q$ at the energy at which the TDE flux is maximum is always greater than 1, implying that most TDEs are produced in the full loss cone regime. Depending on the choice of the IMF, the value of $q$ is $\approx 2-10$ at the beginning of the evolution and it reaches $\approx 10-20$ towards the end, since the maximum flux of TDEs progressively gets to lower binding energies (thus larger radii). It can be shown (see. e.g.~\citealt{Stone2020} and references therein) that the flux in the loss cone scales nearly as $\propto 1/\ln(l_{\rm lc}/l_0)$ if $q\ll1$, and it is thus very little dependent on the size of the loss cone; instead, it scales nearly as  $\propto (l_{\rm lc}/l_0)^2$ if $q\gg1$, so that the loss cone size can have a non-negligible impact on TDE rates. Since in the studied system most TDEs came from the full loss cone, the residual curve shown in Fig.~\ref{fig:rel_rate_mass} resembles the shape of $r_t(m)$.

\subsection{Preference for the disruption of stars with $m>1.3 \msun$}

Recently, \citet{Mockler2021} were able to constrain the  nitrogen-to-carbon abundance ratios for a subset of the observed TDEs. Since only stars initially more massive than $\approx 1.3 \msun$ undergo the CNO cycle, these moderately massive stars are nitrogen-rich and carbon-deficient, and they can be spotted via spectral analysis. Considering the galaxy sample presented by \citet[][]{French2020}, they find that 2 of 5 TDE candidates (40 per cent) in E+A galaxies and 3 of 13 (23 per cent) in quiescent Balmer-strong galaxies have enhanced nitrogen abundances; this is a strict lower limit, since part of the TDEs in the sample do not have C/N measurements. 

\begin{figure}
\centering
\includegraphics[ width=0.38\textwidth]{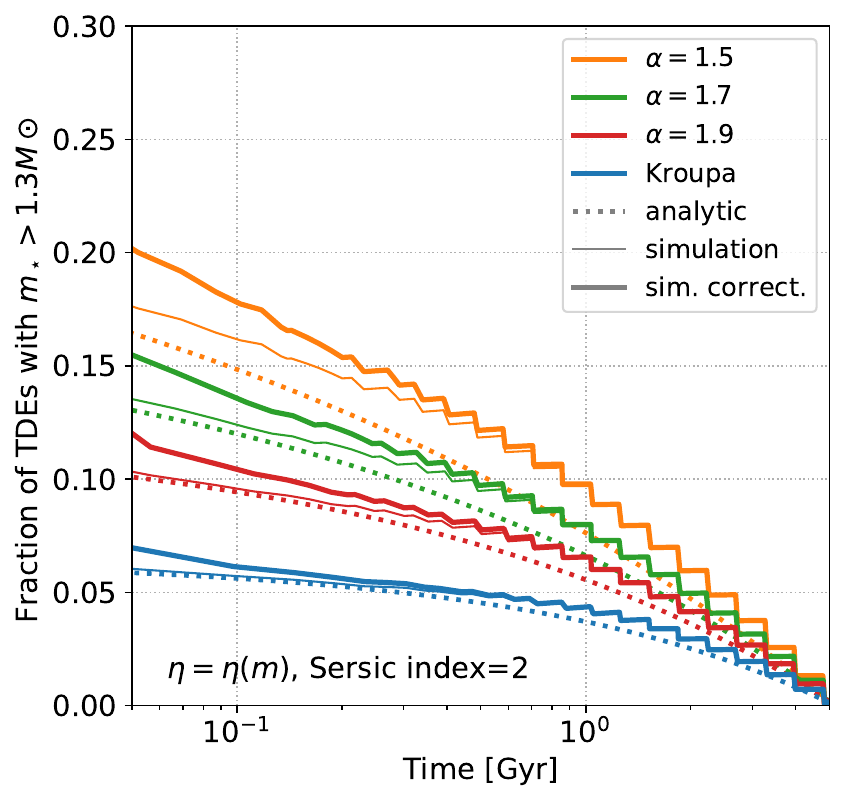}
\includegraphics[ width=0.38\textwidth]{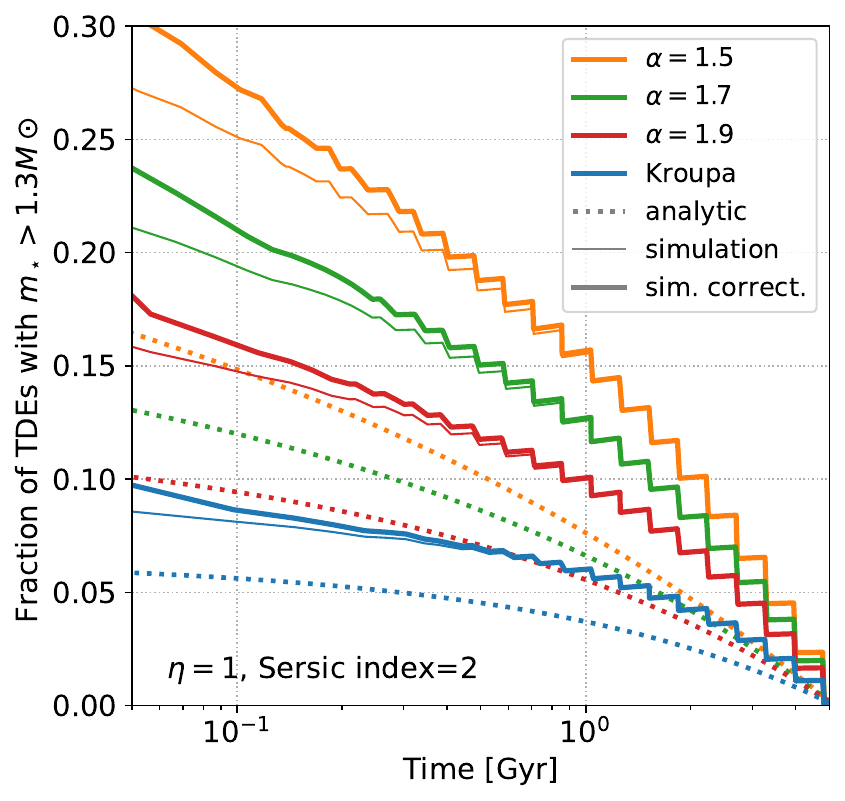}
    \caption{The plots show  the probability of disrupting a star with initial mass above 1.3 $\msun$ (i.e. a star that is expected to exibit a C/N overabundance in the TDE spectrum) as a function of time for different choices of the IMF, as color-coded in the legend. The dotted lines are obtained by a simple estimate based on the IMF and on the time needed by each star to become a WD. The solid thin lines show the results as extracted from the Fokker-Plank runs, while the solid thick lines show the same curves corrected based on the results of Sec.~\ref{sec:mass_preference}, i.e. accounting for the fact that heavy stars -- which are not self consistently simulated -- have larger stellar radii and are thus more likely to be disrupted (see e.g. footnotes~\ref{fn:extrap_eta},\ref{fn:extrap_no_eta}); this thick line should be considered the reference one to make comparisons with the observational data. Both panels show runs with nuclear cluster Sersic index set to 2. The top plot shows results from the default runs with TDE radius defined via Eq.~\ref{eq:eta_ryu}; the bottom one assumes $\eta=1$. The step-like shape of the lines is due to the fact that the simulations feature discrete mass families instead of a continuus stellar mass function.}
    \label{fig:cnover}
\end{figure}

In order to compare our results with their findings, Fig.~\ref{fig:cnover} shows the fraction of TDEs coming from stars more massive than 1.3$\msun$, obtained from an analytical estimate based only on the mass function and age (Eq.~\ref{eq:time_mass}); in addition, we show the same fraction as extracted from the simulations -- with a further large mass correction. Specifically, on the top of the early time mass corrections described in Sec.~\ref{sec:stellar_evolution_corrections}, we further correct for the fact that massive stars are progressively more likely to be destroyed due to their larger radii, according to the $m>2\msun$ extrapolations presented in Sec.~\ref{sec:mass_preference} (computed for all IMFs).

The fraction of relatively massive disrupted stars is very close to the analytical estimate (top panel, dotted) if $r_t$ is defined via Eq.~\ref{eq:eta_ryu}, i.e. $\eta=\eta(m)$; this is due to the fact that in this case massive stars are disrupted nearly as predicted by the IMF, and the ratio in the top panel of Fig.~\ref{fig:rel_rate_mass} is very close to 1. However, if one  fixes  $\eta=1$ (bottom panel, Fig.~\ref{fig:cnover}) the fraction of disrupted stars above $1.3\msun$ gets significantly larger compared with the analytical estimate;  this is due to the fact that if $\eta=1$,  stars above 1.3 $\msun$ are significantly more likely to be disrupted compared with what would be estimated based on the IMF only (bottom panel of Fig.~\ref{fig:rel_rate_mass}).

In general, we can conclude that more top-heavy IMFs can result in up to $\approx15$ (25) per cent of TDEs to come from moderately massive stars over the first hundreds of Myr if $\eta=\eta(m)$ ($\eta=1$). The results by \citet{Mockler2021} suggest that the observed fraction is likely  larger;  however, it is important to keep in mind that currently their sample is very small and thus subject to large statistical errors. Notably, the results also imply that a different prescription for $r_t$ can significantly affect the probability of disrupting moderately massive stars. \footnote{This also implies that assuming  stellar radii to be fixed in time may impact the relative rates as a function of the stellar mass, since stellar radii can vary  non-negligibly along the main sequence \citep[e.g.][]{Bressan2012}.}

\subsection{TDE rates in case of an extended central star formation}

\begin{figure}
\centering
\includegraphics[ width=0.38\textwidth]{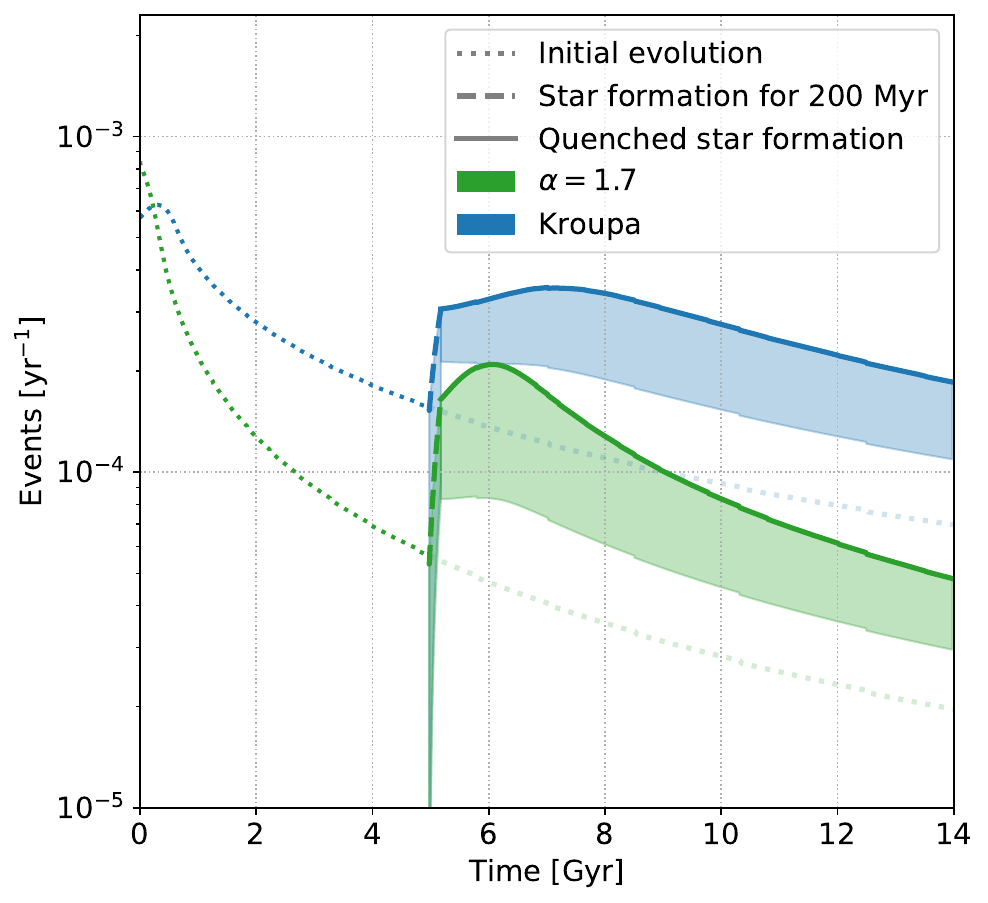}
    \caption{TDE rates for the Kroupa (blue) and top-heavy, $\alpha=1.7$ case (green) if star formation is switched on at $t=5$ Gyr, grows the mass of the central cluster by 30 per cent and is quenched after 200 Myr. The TDE rates before 5 Gyr (dotted lines) are the same shown in Fig.~\ref{fig:time_segregation}; the dashed lines show the event rate during star formation, and the solid lines show it after star formation is quenched. In all cases, the shown rate comes from the full stellar population;  after star formation, the lower edge of the shaded area shows the event rate coming from the former population as a function of time: this alone also grows as a result of star formation.}
    \label{fig:star_formation}
\end{figure}

In order to better constrain the effect of star formation, instead of assuming the entire stellar cluster to be formed instantaneously at $t=0$,  here we test the effect of an extended star formation prescription. We take the snapshot at $t=5$ Gyr of the nuclear cluster simulations described above (whose rates are shown in Fig.~\ref{fig:time_segregation}), and in particular we refer to the cases with a Kroupa and a top-heavy, $\alpha=1.7$ IMF, both with Sersic index equal to 3. We switch on star formation in the inner $\sim 6$ pc for $200$ Myr,  growing the central cluster mass by 30 per cent, and then we quench star formation (see Sec.~\ref{sec:galaxy} for more technical details). The obtained rates are shown in Fig.~\ref{fig:star_formation}. Even if only 23 per cent of the new mass of the cluster comes from the recent star formation, the TDE rates grow by a factor of a few ($\times 2-4$)  compared to the values right before it started; the impact is more dramatic in the case of a top-heavy IMF. Not only the event rate is boosted by TDEs coming from newly formed stars, but also the former population gets its rate enhanced as star formation occurs, as shown in Fig.~\ref{fig:star_formation}.
The fraction of TDEs coming from the newly formed stars with respect to the total rate  for the top-heavy (Kroupa) case reaches a maximum of about 60 (45) per cent near the peak after star formation, and declines to nearly 39 (41) per cent by the end of the evolution. 

The TDE rate boost necessarily depends on the choice of the fraction of newly formed stars, on where they form and so on. Still,  the presented results clearly show that even assuming only a fraction of the cluster to be formed in the starburst leads to a significant growth in TDE rates,  as the perturbed former population has its  rates enhanced while on its way to a new equilibrium state.

\section{Summary and discussion}\label{sec:discussion}

This paper explores the effect of a complete, evolved stellar mass function on the rate of TDEs in a nucleated galaxy analogous to observed TDE hosts.  We study the event rates in systems (i) featuring a Kroupa and a set of top-heavy IMFs, (ii) featuring different central cluster concentrations (iii) assuming an instantaneous and a more extended star formation.  Below we summarize our key results.
\begin{enumerate}
    \item The bulge virtually does not contribute to TDE rates.
    \item Assuming that the nuclear cluster is present and it instantaneously forms at $t=0$, we find that the TDE rate drops significantly (by a factor $\sim 10-100$, Fig.~\ref{fig:time_segregation}) over 14 Gyr if one accounts for a complete mass function, while the drop is only a factor of a few if we consider a population of 1 $\msun$ stars. The rate drop  is more pronounced if one starts with an initially more concentrated stellar cluster, and if the IMF is more top-heavy. The ratio between the early and the late event rates (Tab.~\ref{tab:time_enhancements_nsc}) can explain the overabundance of post-starburst galaxies among TDE hosts, especially if a  top-heavy IMF is assumed and/or the initial density profile of the central cluster is more concentrated.
    \item The initial peak in the event rates starts declining after a mass segregation  time-scale. Within this timeframe, both mass segregation occurs and a \citeauthor{Bahcall1976} cusp develops near the MBH (Fig.~\ref{fig:density_evolution}). Afterwards, the event rate declines as a result of the stellar system expansion induced by two-body relaxation, that brings to a drop in the density normalization, while the radius from which most TDEs are produced grows larger.
    \item The fraction of TDEs coming from stars of different mass can be described by the stellar mass function to the 0-th order. However, most of the destroyed stars come from the full loss cone regime, for which the event rate scales nearly linearly with the tidal disruption radius $r_t(m)$, which grows  with the stellar mass $m$; it follows that there is a residual modulation in the fraction of destroyed stars as a function of $m$, that resembles the trend of $r_t(m)$ (Fig.~\ref{fig:rel_rate_mass}). Depending on the definition of $r_t(m)$, the number of destroyed stars as a function of their mass can vary significantly. 
    \item We find that,  a few hundreds of Myr after the starburst, up to $\approx 15-25$ per cent of destroyed stars have  $m>1.3\msun$  (Fig.~\ref{fig:cnover}). This fraction  is lower compared with what has been observationally constrained by \citet[][]{Mockler2021}, but still relatively close to it; varying the definition of $r_t(m)$  significantly affects this estimate.
    \item If we assume only 30 per cent of the cluster  mass  to be formed in a 200-Myr star formation episode (happening when the older cluster stellar population is already relaxed), the TDE rate   grows by a factor of a few after the starburst. A sensible part of the enhancement comes from the older stellar population, whose equilibrium state is perturbed by the freshly injected mass.
\end{enumerate}

\begin{figure}
\centering
\includegraphics[ width=0.42\textwidth]{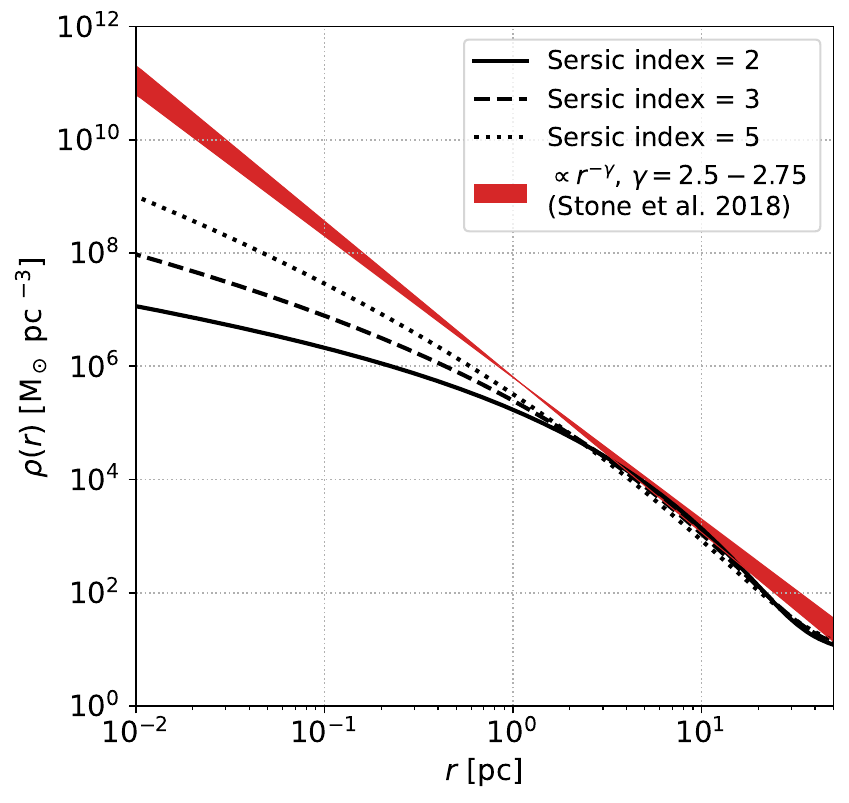}
    \caption{Comparison between our initial conditions based on a Milky-Way like model (black lines)   with those employed by \citet{Stone2018}. They explore a range of cusps $\rho\propto r^{-\gamma}$, where $\gamma=1.75-2.75$ , and they conclude that the post-starburst overrepresentation in their monochromatic runs can be explained if $\gamma>2.5$ (red area); their density normalization has been set via $\rho(1$ pc$)=0.617^\gamma \times10^{6.32}\msun\, {\rm pc}^{-3} (M_\bullet/10^6\msun)^{-1.05+0.425\gamma}$ (see their footnote 12). The density profiles that justify the post-starburst preference in \citet{Stone2018}  are several orders of magnitude denser in the inner regions compared with our models.}
    \label{fig:compare_ic}
\end{figure}

\begin{figure}
\centering
\includegraphics[ width=0.42\textwidth]{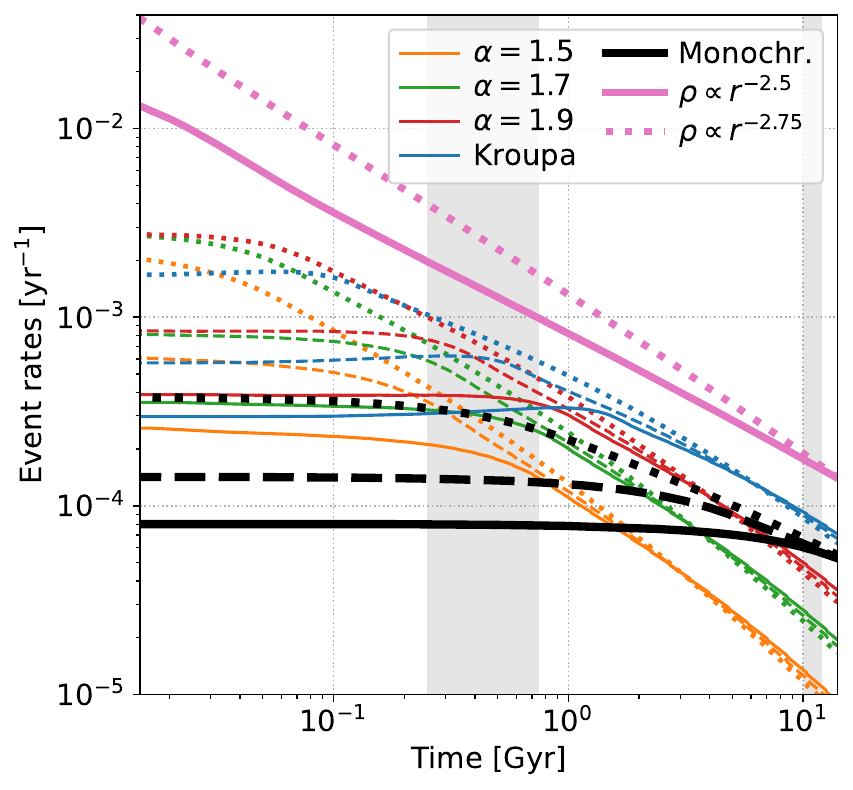}
    \caption{Comparison between the TDE rates obtained by \citet{Stone2018} with the ultrasteep cusp (pink, thick lines featuring the highest rates;  solid line: $\rho\propto r^{-2.5}$; dotted line: $\rho\propto r^{-2.75}$) and our TDE rates (our rates are also shown in Fig.~\ref{fig:time_segregation}, where it is easier to distinguish each different line). The different IMFs adopted in our runs are color-coded as in the legend.  The solid, dashed and dotted lines refer to our models with Sersic index respectively equal to 2, 3, 5. The shaded regions mark where the average early (250$-$750 Myr) and late ($10-12$ Gyr) TDE rates are computed in Tab.~\ref{tab:time_enhancements_nsc} -- numbers in parenthesis. For comparison, in the same regions, the TDE enhancements for the \citet[][]{Stone2018} models are respectively equal to 8, 14 if $\gamma = 2.5, 2.75$. }
    \label{fig:stone_comparison}
\end{figure}

One of our key results  is  that accounting for a complete stellar mass function can explain the post-starburst preference of TDE hosts. The alternative, currently most widespread explanation is that post-starburst galaxies are initially born with an extremely steep density profile \citep[][]{Stone2018}. Fig.~\ref{fig:compare_ic} compares the initial conditions adopted in our runs with the ultrasteep initial conditions adopted by \citet[][]{Stone2018} to explain the post-starburst preference, rescaled to our MBH mass according to their footnote~12 (see also the caption of our  Fig.~\ref{fig:compare_ic}). The central density in our and their profile differs by several orders of magnitude, and in general it is not straightforward to imagine how star formation could generate such an extremely steep cusp. In fact, \citet[][]{Sanders1998} points out that, too close to the MBH, molecular clouds should  get disrupted by its gravitational field preventing star formation; even if the Milky Way nucleus features a population of young stars very close to the MBH \citep[][]{Morris1993,Ghez2003, Lu2013}, supporting instead the idea of possible central star formation, it seems unlikely it can develop such steep profiles. For instance, in these systems the time-scale for stellar collisions and mergers can be shorter than the relaxation time-scale at small radii ($<10^{-2}$ pc, \citealt{Stone2018}).
Note that if the density profile is initially very steep, relaxation processes  render it milder in time, as the equilibrium configuration is the \citet{Bahcall1976} solution ($\rho\propto r^{-1.75}$): this means that the ultrasteep configuration should be in place since the starburst stage, i.e. stars should be basically born with $\rho\propto r^{-\gamma}, \gamma=2.5-2.75$. 
In order to compare our results with the ones by \citet[][]{Stone2018}, we evolve their (monochromatic) profiles shown in Fig.~\ref{fig:compare_ic} with $\rho \propto r^{-2.5}, r^{-2.75}$ and we compare the obtained rates with our complete mass function models in Fig.~\ref{fig:stone_comparison}. Even if their  rates are overall larger, the early to late event rate is compatible with what we find for much milder profiles featuring a complete mass function; in fact, the ratio between the average TDE rate in the range 250$-$750 Myr (i.e. roughly the age of post-starburst systems)  and $10-12$ Gyr is equal to $8-14$ for the \citet[][]{Stone2018} cases, i.e. they are compatible with or smaller than the same ratios computed in our runs (see numbers in parenthesis in Tab.~\ref{tab:time_enhancements_nsc}). 

It is also  worth stressing that the two-body relaxation driven expansion that results in a more or less prompt rate drop acts in both the complete mass function and  monochromatic case; in fact, it ultimately represents the driver of the rate decline in the  ultra-steep cusps of \citet{Stone2018}. 
Expansion starts later and is less efficient in the monochromatic cases, but  Fig.~\ref{fig:time_segregation} clearly shows that the TDE rates start  declining in a nearly power-law fashion at late times even in the single-mass runs. Since lower mass MBHs  are likely to be embedded in systems with shorter relaxation times, it is reasonable to expect that these systems would relax over a time-scale significantly shorter than a Hubble time even in the monochromatic case, and thus start expanding and initiate their rate decline much earlier. As a consequence, in systems featuring low mass MBHs, the ratio between the early and late TDE rate could be larger than what found here for the single mass scenario. Nonetheless, a complete mass function would likely enhance the rate drop even about lower mass MBHs; in addition, we stress that the post-starburst galaxies that host observed TDEs harbour  an  MBH whose mass estimate  is compatible with the one  chosen in this  work \citep[][]{French2020}. \\

Our results  support the  idea  that starbursting systems are likely characterized by
at least a moderately top-heavy IMF, as already suggested in the literature \citep{Zhang2018}. In fact, our computed early to late event rates  are similar to the observations-based post-starburst overrepresentation ($15-30$) if  $\alpha\lesssim1.7$. Furthermore, a top-heavy IMF can help explain the large fraction of moderately massive disrupted stars in post-starburst galaxies  \citep[][Fig.~\ref{fig:cnover}]{Mockler2021}. Interestingly, this implies that in principle TDE observations can be adopted to constrain the IMF of starbursting galaxies as more TDE observations become available.
Another interesting point concerns the choice of the prescription for computing the tidal disruption radius $r_t=\eta (M_\bullet/m)^{1/3} R_\star$. Changing the prescription of $\eta$, the probability of disrupting stars with different masses changes significantly. This means that more work is needed to best constrain the functional form of $\eta$, as our results suggest that e.g. accounting for partial TDEs can affect the relative probability of disrupting more or less massive stars. 

Concerning the overall estimate of TDE rates, theoretical studies (typically assuming a monochromatic mass function) tend to predict  values that are generally  larger than the observed ones (respectively a few $\times10^{-4}$ versus $0.3-1\times 10^{-4}$ events per galaxy per year, see e.g. \citealt{French2020, Stone2016}); this discrepancy is worsened by the post-starburst overrepresentation, that makes observed rates for quiescent galaxies lower by a factor $\sim 3$. Interestingly enough, the ratio between the late event rates of a monochromatic and a complete mass function $\dot{N}_{\rm mc}/\dot{N}_{\rm MF}$ varies significantly depending the chosen IMF.
In the Kroupa case, this ratio is very close to 1 ($\dot{N}_{\rm mc}/\dot{N}_{\rm MF}=0.75-0.86$); however,  choosing a top-heavier IMF makes the late time event rates much smaller than their corresponding value for a monochromatic mass function. In fact, $\dot{N}_{\rm mc}/\dot{N}_{\rm MF}\approx 6,3,1.5$  if $\alpha=1.5, 1.7, 1.9$ implying that assuming, in general, a top heavy IMF to be in place in galactic nuclei would help to reconcile the observed and detected TDE rates. This obviously would work only if all galaxies dominating TDE rates (e.g. more concentrated ones, featuring a nuclear star cluster and a relatively low MBH mass) are characterized by a top-heavy IMF. To better constrain this aspect,  it would  be important to adopt  an observationally motivated galaxy and MBH sample, in order to assess in more detail the statistics of TDE rates when assuming  different complete mass functions. This will be the subject of a future work.

To conclude, it is worth mentioning that one  important assumption introduced  in most of our runs is that the whole galactic nucleus is formed  at $t=0$. It may seem unrealistic the whole nuclear cluster is formed instantaneously; however, this assumption would be compatible with the mass fraction generated by the starburst in a post-starburst galaxy (Sec.~\ref{sec:intro}, also note that the same assumption of instantaneous formation is made in \citealt[][]{Stone2018}). The main process that leads to the formation of nuclear star clusters is still a debated topic, and there are two main proposed formation channels: in the 
in-situ (wet)  scenario, the nuclear cluster is generated via efficient star formation occurring at the galaxy centre \citep[e.g.][]{Loose1982}; in the migration scenario, stellar clusters formed elsewhere migrate via dynamical friction in the galaxy centre, to form a nuclear overdensity \citep[e.g.][]{Tremaine1975, Capuzzo-Dolcetta1993}.\footnote{More recently, a combined, wet migration scenario has also been proposed \citep[][]{Guillard2016}.} In the  in-situ framweork, the instantaneous approximation for the cluster formation is reasonable as the star formation time-scale in post-starburst systems can be as low as a few tens of Myr \citep{French2018}, much shorter than the typical galaxy evolution time-scales. Even in the migration scenario, the sudden perturbation in the potential caused by a cluster inspiral would result in a strong dynamical perturbation for the central orbits, so that it is reasonable to assume  the system instantaneously forms at that  moment. More in general, 
since mass segregation is in fact the main trigger of the burst in TDE rates, any  event violent enough to strongly perturb the central galaxy potential well and quash the previously established, mass-segregated cusp would likely  result in a strong rate enhancement.  
Phenomena like a significant inflow of gas possibly resulting in a strong  burst of star formation, a galaxy merger \citep{Gualandris2012jan}, the inspiral of a massive cluster near the nucleus \citep{Bortolas2018, Arca-Sedda2017}, could thus induce a TDE burst. Addressing in detail these aspects would require different strategies than those adopted in the present work, and we plan to explore this in a forthcoming study.

\section*{Acknowledgements}

We warmly thank  the anonymous referee for their very useful comments and suggestions.
EB warmly thanks Matteo Bonetti, Luca Broggi,  Alberto Sesana and Nicholas Stone for fruitful discussion, and Eugene Vasiliev for the support in the use of Phaseflow.
EB acknowledges support from the European Research Council (ERC) under the European Union's Horizon 2020 research and innovation program ERC-2018-COG
under grant agreement N.~818691 (B~Massive). EB also acknowledges support from the University of Milano-Bicocca through the `Giovani Talenti' prize.

\section*{Data Availability Statement}
The data underlying this article will be shared on reasonable request to the corresponding author.




\bibliography{bibliography} 

\begin{thebibliography}{}
\makeatletter
\relax
\def\mn@urlcharsother{\let\do\@makeother \do\$\do\&\do\#\do\^\do\_\do\%\do\~}
\def\mn@doi{\begingroup\mn@urlcharsother \@ifnextchar [ {\mn@doi@}
  {\mn@doi@[]}}
\def\mn@doi@[#1]#2{\def\@tempa{#1}\ifx\@tempa\@empty \href
  {http://dx.doi.org/#2} {doi:#2}\else \href {http://dx.doi.org/#2} {#1}\fi
  \endgroup}
\def\mn@eprint#1#2{\mn@eprint@#1:#2::\@nil}
\def\mn@eprint@arXiv#1{\href {http://arxiv.org/abs/#1} {{\tt arXiv:#1}}}
\def\mn@eprint@dblp#1{\href {http://dblp.uni-trier.de/rec/bibtex/#1.xml}
  {dblp:#1}}
\def\mn@eprint@#1:#2:#3:#4\@nil{\def\@tempa {#1}\def\@tempb {#2}\def\@tempc
  {#3}\ifx \@tempc \@empty \let \@tempc \@tempb \let \@tempb \@tempa \fi \ifx
  \@tempb \@empty \def\@tempb {arXiv}\fi \@ifundefined
  {mn@eprint@\@tempb}{\@tempb:\@tempc}{\expandafter \expandafter \csname
  mn@eprint@\@tempb\endcsname \expandafter{\@tempc}}}

\bibitem[\protect\citeauthoryear{{Alexander}}{{Alexander}}{2017}]{Alexander2017}
{Alexander} T.,  2017, in Journal of Physics Conference Series. p. 012019
  (\mn@eprint {arXiv} {1702.00597}), \mn@doi{10.1088/1742-6596/840/1/012019}

\bibitem[\protect\citeauthoryear{{Alexander} \& {Hopman}}{{Alexander} \&
  {Hopman}}{2009}]{Alexander2009}
{Alexander} T.,  {Hopman} C.,  2009, \mn@doi [\apj]
  {10.1088/0004-637X/697/2/1861}, \href
  {http://adsabs.harvard.edu/abs/2009ApJ...697.1861A} {697, 1861}

\bibitem[\protect\citeauthoryear{{Amaro-Seoane} \& {Preto}}{{Amaro-Seoane} \&
  {Preto}}{2011}]{Amaro-Seoane2011}
{Amaro-Seoane} P.,  {Preto} M.,  2011, \mn@doi [Classical and Quantum Gravity]
  {10.1088/0264-9381/28/9/094017}, \href
  {http://adsabs.harvard.edu/abs/2011CQGra..28i4017A} {28, 094017}

\bibitem[\protect\citeauthoryear{{Arca-Sedda}, {Berczik}, {Capuzzo-Dolcetta},
  {Fragione}, {Sobolenko}  \& {Spurzem}}{{Arca-Sedda}
  et~al.}{2017}]{Arca-Sedda2017}
{Arca-Sedda} M.,  {Berczik} P.,  {Capuzzo-Dolcetta} R.,  {Fragione} G.,
  {Sobolenko} M.,   {Spurzem} R.,  2017, preprint, \href
  {http://adsabs.harvard.edu/abs/2017arXiv171205810A} {} (\mn@eprint {arXiv}
  {1712.05810})

\bibitem[\protect\citeauthoryear{{Arcavi} et~al.,}{{Arcavi}
  et~al.}{2014}]{Arcavi2014}
{Arcavi} I.,  et~al., 2014, \mn@doi [\apj] {10.1088/0004-637X/793/1/38}, \href
  {https://ui.adsabs.harvard.edu/abs/2014ApJ...793...38A} {793, 38}

\bibitem[\protect\citeauthoryear{{Bahcall} \& {Wolf}}{{Bahcall} \&
  {Wolf}}{1976}]{Bahcall1976}
{Bahcall} J.~N.,  {Wolf} R.~A.,  1976, \mn@doi [\apj] {10.1086/154711}, \href
  {http://adsabs.harvard.edu/abs/1976ApJ...209..214B} {209, 214}

\bibitem[\protect\citeauthoryear{{Bahcall} \& {Wolf}}{{Bahcall} \&
  {Wolf}}{1977}]{Bahcall1977}
{Bahcall} J.~N.,  {Wolf} R.~A.,  1977, \mn@doi [\apj] {10.1086/155534}, \href
  {http://adsabs.harvard.edu/abs/1977ApJ...216..883B} {216, 883}

\bibitem[\protect\citeauthoryear{{Barcons} et~al.,}{{Barcons}
  et~al.}{2012}]{Barcons2012}
{Barcons} X.,  et~al., 2012, arXiv e-prints, \href
  {https://ui.adsabs.harvard.edu/abs/2012arXiv1207.2745B/abstract} {p.
  arXiv:1207.2745}

\bibitem[\protect\citeauthoryear{{Bartko} et~al.,}{{Bartko}
  et~al.}{2010}]{Bartko2010}
{Bartko} H.,  et~al., 2010, \mn@doi [\apj] {10.1088/0004-637X/708/1/834}, \href
  {https://ui.adsabs.harvard.edu/abs/2010ApJ...708..834B} {708, 834}

\bibitem[\protect\citeauthoryear{{Bellm} et~al.,}{{Bellm}
  et~al.}{2019}]{Bellm2019}
{Bellm} E.~C.,  et~al., 2019, \mn@doi [\pasp] {10.1088/1538-3873/aaecbe}, \href
  {https://ui.adsabs.harvard.edu/abs/2019PASP..131a8002B} {131, 018002}

\bibitem[\protect\citeauthoryear{{Bortolas} \& {Mapelli}}{{Bortolas} \&
  {Mapelli}}{2019}]{Bortolas2019}
{Bortolas} E.,  {Mapelli} M.,  2019, \mn@doi [\mnras] {10.1093/mnras/stz440},
  \href {https://ui.adsabs.harvard.edu/abs/2019MNRAS.485.2125B} {485, 2125}

\bibitem[\protect\citeauthoryear{{Bortolas}, {Mapelli}  \& {Spera}}{{Bortolas}
  et~al.}{2017}]{Bortolas2017}
{Bortolas} E.,  {Mapelli} M.,   {Spera} M.,  2017, \mn@doi [\mnras]
  {10.1093/mnras/stx930}, \href
  {http://adsabs.harvard.edu/abs/2017MNRAS.469.1510B} {469, 1510}

\bibitem[\protect\citeauthoryear{{Bortolas}, {Mapelli}  \& {Spera}}{{Bortolas}
  et~al.}{2018}]{Bortolas2018}
{Bortolas} E.,  {Mapelli} M.,   {Spera} M.,  2018, \mn@doi [\mnras]
  {10.1093/mnras/stx2795}, \href
  {http://adsabs.harvard.edu/abs/2018MNRAS.474.1054B} {474, 1054}

\bibitem[\protect\citeauthoryear{{Bressan}, {Marigo}, {Girardi}, {Salasnich},
  {Dal Cero}, {Rubele}  \& {Nanni}}{{Bressan} et~al.}{2012}]{Bressan2012}
{Bressan} A.,  {Marigo} P.,  {Girardi} L.,  {Salasnich} B.,  {Dal Cero} C.,
  {Rubele} S.,   {Nanni} A.,  2012, \mn@doi [\mnras]
  {10.1111/j.1365-2966.2012.21948.x}, \href
  {http://adsabs.harvard.edu/abs/2012MNRAS.427..127B} {427, 127}

\bibitem[\protect\citeauthoryear{{Capuzzo-Dolcetta}}{{Capuzzo-Dolcetta}}{1993}]{Capuzzo-Dolcetta1993}
{Capuzzo-Dolcetta} R.,  1993, \mn@doi [\apj] {10.1086/173189}, \href
  {http://adsabs.harvard.edu/abs/1993ApJ...415..616C} {415, 616}

\bibitem[\protect\citeauthoryear{{Chambers} et~al.,}{{Chambers}
  et~al.}{2016}]{Chambers2016}
{Chambers} K.~C.,  et~al., 2016, arXiv e-prints, \href
  {https://ui.adsabs.harvard.edu/abs/2016arXiv161205560C} {p. arXiv:1612.05560}

\bibitem[\protect\citeauthoryear{{Chen}, {Girardi}, {Bressan}, {Marigo},
  {Barbieri}  \& {Kong}}{{Chen} et~al.}{2014}]{Chen2014nov}
{Chen} Y.,  {Girardi} L.,  {Bressan} A.,  {Marigo} P.,  {Barbieri} M.,   {Kong}
  X.,  2014, \mn@doi [\mnras] {10.1093/mnras/stu1605}, \href
  {http://adsabs.harvard.edu/abs/2014MNRAS.444.2525C} {444, 2525}

\bibitem[\protect\citeauthoryear{{Cummings}, {Kalirai}, {Tremblay},
  {Ramirez-Ruiz}  \& {Choi}}{{Cummings} et~al.}{2018}]{Cummings2018}
{Cummings} J.~D.,  {Kalirai} J.~S.,  {Tremblay} P.~E.,  {Ramirez-Ruiz} E.,
  {Choi} J.,  2018, \mn@doi [\apj] {10.3847/1538-4357/aadfd6}, \href
  {https://ui.adsabs.harvard.edu/abs/2018ApJ...866...21C} {866, 21}

\bibitem[\protect\citeauthoryear{{D'Orazio}, {Loeb}  \&
  {Guillochon}}{{D'Orazio} et~al.}{2019}]{D'Orazio2019}
{D'Orazio} D.~J.,  {Loeb} A.,   {Guillochon} J.,  2019, \mn@doi [\mnras]
  {10.1093/mnras/stz652}, \href
  {https://ui.adsabs.harvard.edu/abs/2019MNRAS.485.4413D} {485, 4413}

\bibitem[\protect\citeauthoryear{{Davis}, {Graham}  \& {Cameron}}{{Davis}
  et~al.}{2019}]{Davis2019}
{Davis} B.~L.,  {Graham} A.~W.,   {Cameron} E.,  2019, \mn@doi [\apj]
  {10.3847/1538-4357/aaf3b8}, \href
  {https://ui.adsabs.harvard.edu/abs/2019ApJ...873...85D} {873, 85}

\bibitem[\protect\citeauthoryear{{Emami} \& {Loeb}}{{Emami} \&
  {Loeb}}{2020}]{Emami2020}
{Emami} R.,  {Loeb} A.,  2020, \mn@doi [\jcap] {10.1088/1475-7516/2020/02/021},
  \href {https://ui.adsabs.harvard.edu/abs/2020JCAP...02..021E} {2020, 021}

\bibitem[\protect\citeauthoryear{{Foote}, {Generozov}  \& {Madigan}}{{Foote}
  et~al.}{2020}]{Foote2020}
{Foote} H.~R.,  {Generozov} A.,   {Madigan} A.-M.,  2020, \mn@doi [\apj]
  {10.3847/1538-4357/ab6c66}, \href
  {https://ui.adsabs.harvard.edu/abs/2020ApJ...890..175F} {890, 175}

\bibitem[\protect\citeauthoryear{{French}}{{French}}{2021}]{French2021}
{French} K.~D.,  2021, \mn@doi [\pasp] {10.1088/1538-3873/ac0a59}, \href
  {https://ui.adsabs.harvard.edu/abs/2021PASP..133g2001F} {133, 072001}

\bibitem[\protect\citeauthoryear{{French}, {Arcavi}  \& {Zabludoff}}{{French}
  et~al.}{2016}]{French2016}
{French} K.~D.,  {Arcavi} I.,   {Zabludoff} A.,  2016, \mn@doi [\apjl]
  {10.3847/2041-8205/818/1/L21}, \href
  {https://ui.adsabs.harvard.edu/abs/2016ApJ...818L..21F} {818, L21}

\bibitem[\protect\citeauthoryear{{French}, {Arcavi}  \& {Zabludoff}}{{French}
  et~al.}{2017}]{French2017}
{French} K.~D.,  {Arcavi} I.,   {Zabludoff} A.,  2017, \mn@doi [\apj]
  {10.3847/1538-4357/835/2/176}, \href
  {https://ui.adsabs.harvard.edu/abs/2017ApJ...835..176F} {835, 176}

\bibitem[\protect\citeauthoryear{{French}, {Yang}, {Zabludoff}  \&
  {Tremonti}}{{French} et~al.}{2018}]{French2018}
{French} K.~D.,  {Yang} Y.,  {Zabludoff} A.~I.,   {Tremonti} C.~A.,  2018,
  \mn@doi [\apj] {10.3847/1538-4357/aacb2d}, \href
  {https://ui.adsabs.harvard.edu/abs/2018ApJ...862....2F} {862, 2}

\bibitem[\protect\citeauthoryear{{French}, {Wevers}, {Law-Smith}, {Graur}  \&
  {Zabludoff}}{{French} et~al.}{2020}]{French2020}
{French} K.~D.,  {Wevers} T.,  {Law-Smith} J.,  {Graur} O.,   {Zabludoff}
  A.~I.,  2020, \mn@doi [\ssr] {10.1007/s11214-020-00657-y}, \href
  {https://ui.adsabs.harvard.edu/abs/2020SSRv..216...32F} {216, 32}

\bibitem[\protect\citeauthoryear{{Generozov}, {Stone}, {Metzger}  \&
  {Ostriker}}{{Generozov} et~al.}{2018}]{Generozov2018}
{Generozov} A.,  {Stone} N.~C.,  {Metzger} B.~D.,   {Ostriker} J.~P.,  2018,
  \mn@doi [\mnras] {10.1093/mnras/sty1262}, \href
  {https://ui.adsabs.harvard.edu/abs/2018MNRAS.478.4030G} {478, 4030}

\bibitem[\protect\citeauthoryear{{Ghez} et~al.,}{{Ghez}
  et~al.}{2003}]{Ghez2003}
{Ghez} A.~M.,  et~al., 2003, \mn@doi [\apjl] {10.1086/374804}, \href
  {http://adsabs.harvard.edu/abs/2003ApJ...586L.127G} {586, L127}

\bibitem[\protect\citeauthoryear{{Graur}, {French}, {Zahid}, {Guillochon},
  {Mandel}, {Auchettl}  \& {Zabludoff}}{{Graur} et~al.}{2018}]{Graur2018}
{Graur} O.,  {French} K.~D.,  {Zahid} H.~J.,  {Guillochon} J.,  {Mandel} K.~S.,
   {Auchettl} K.,   {Zabludoff} A.~I.,  2018, \mn@doi [\apj]
  {10.3847/1538-4357/aaa3fd}, \href
  {https://ui.adsabs.harvard.edu/abs/2018ApJ...853...39G} {853, 39}

\bibitem[\protect\citeauthoryear{{Gravity Collaboration} et~al.,}{{Gravity
  Collaboration} et~al.}{2020}]{Gravity2020}
{Gravity Collaboration} et~al., 2020, \mn@doi [\aap]
  {10.1051/0004-6361/202037813}, \href
  {https://ui.adsabs.harvard.edu/abs/2020A&A...636L...5G} {636, L5}

\bibitem[\protect\citeauthoryear{{Gualandris} \& {Merritt}}{{Gualandris} \&
  {Merritt}}{2012}]{Gualandris2012jan}
{Gualandris} A.,  {Merritt} D.,  2012, \mn@doi [\apj]
  {10.1088/0004-637X/744/1/74}, \href
  {http://adsabs.harvard.edu/abs/2012ApJ...744...74G} {744, 74}

\bibitem[\protect\citeauthoryear{{Guillard}, {Emsellem}  \&
  {Renaud}}{{Guillard} et~al.}{2016}]{Guillard2016}
{Guillard} N.,  {Emsellem} E.,   {Renaud} F.,  2016, \mn@doi [\mnras]
  {10.1093/mnras/stw1570}, \href
  {https://ui.adsabs.harvard.edu/abs/2016MNRAS.461.3620G} {461, 3620}

\bibitem[\protect\citeauthoryear{{Hammerstein} et~al.,}{{Hammerstein}
  et~al.}{2021}]{Hammerstein2021}
{Hammerstein} E.,  et~al., 2021, \mn@doi [\apjl] {10.3847/2041-8213/abdcb4},
  \href {https://ui.adsabs.harvard.edu/abs/2021ApJ...908L..20H} {908, L20}

\bibitem[\protect\citeauthoryear{{Hopkins} \& {Quataert}}{{Hopkins} \&
  {Quataert}}{2010}]{Hopkins2010}
{Hopkins} P.~F.,  {Quataert} E.,  2010, \mn@doi [\mnras]
  {10.1111/j.1365-2966.2010.17064.x}, \href
  {https://ui.adsabs.harvard.edu/abs/2010MNRAS.407.1529H} {407, 1529}

\bibitem[\protect\citeauthoryear{{Hoyer}, {Neumayer}, {Georgiev}, {Seth}  \&
  {Greene}}{{Hoyer} et~al.}{2021}]{Hoyer2021}
{Hoyer} N.,  {Neumayer} N.,  {Georgiev} I.~Y.,  {Seth} A.~C.,   {Greene} J.~E.,
   2021, \mn@doi [\mnras] {10.1093/mnras/stab2277}, \href
  {https://ui.adsabs.harvard.edu/abs/2021MNRAS.507.3246H} {507, 3246}

\bibitem[\protect\citeauthoryear{{Ivezi{\'c}} et~al.,}{{Ivezi{\'c}}
  et~al.}{2019}]{Ivezic2019}
{Ivezi{\'c}} {\v Z}.,  et~al., 2019, \mn@doi [\apj] {10.3847/1538-4357/ab042c},
  \href {http://adsabs.harvard.edu/abs/2019ApJ...873..111I} {873, 111}

\bibitem[\protect\citeauthoryear{{Jansen} et~al.,}{{Jansen}
  et~al.}{2001}]{Jansen2001}
{Jansen} F.,  et~al., 2001, \mn@doi [\aap] {10.1051/0004-6361:20000036}, \href
  {https://ui.adsabs.harvard.edu/abs/2001A&A...365L...1J} {365, L1}

\bibitem[\protect\citeauthoryear{{Kochanek}}{{Kochanek}}{2016}]{Kochanek2016}
{Kochanek} C.~S.,  2016, \mn@doi [\mnras] {10.1093/mnras/stw1290}, \href
  {https://ui.adsabs.harvard.edu/abs/2016MNRAS.461..371K} {461, 371}

\bibitem[\protect\citeauthoryear{{Kroupa}}{{Kroupa}}{2001}]{Kroupa2001}
{Kroupa} P.,  2001, \mn@doi [\mnras] {10.1046/j.1365-8711.2001.04022.x}, \href
  {http://adsabs.harvard.edu/abs/2001MNRAS.322..231K} {322, 231}

\bibitem[\protect\citeauthoryear{{Law-Smith}, {Ramirez-Ruiz}, {Ellison}  \&
  {Foley}}{{Law-Smith} et~al.}{2017}]{Law-Smith2017}
{Law-Smith} J.,  {Ramirez-Ruiz} E.,  {Ellison} S.~L.,   {Foley} R.~J.,  2017,
  \mn@doi [\apj] {10.3847/1538-4357/aa94c7}, \href
  {https://ui.adsabs.harvard.edu/abs/2017ApJ...850...22L} {850, 22}

\bibitem[\protect\citeauthoryear{{Licquia} \& {Newman}}{{Licquia} \&
  {Newman}}{2015}]{Licquia2015}
{Licquia} T.~C.,  {Newman} J.~A.,  2015, \mn@doi [\apj]
  {10.1088/0004-637X/806/1/96}, \href
  {https://ui.adsabs.harvard.edu/abs/2015ApJ...806...96L} {806, 96}

\bibitem[\protect\citeauthoryear{{Lodato} \& {Rossi}}{{Lodato} \&
  {Rossi}}{2011}]{Lodato2011}
{Lodato} G.,  {Rossi} E.~M.,  2011, \mn@doi [\mnras]
  {10.1111/j.1365-2966.2010.17448.x}, \href
  {https://ui.adsabs.harvard.edu/abs/2011MNRAS.410..359L} {410, 359}

\bibitem[\protect\citeauthoryear{{Lodato}, {King}  \& {Pringle}}{{Lodato}
  et~al.}{2009}]{Lodato2009}
{Lodato} G.,  {King} A.~R.,   {Pringle} J.~E.,  2009, \mn@doi [\mnras]
  {10.1111/j.1365-2966.2008.14049.x}, \href
  {https://ui.adsabs.harvard.edu/abs/2009MNRAS.392..332L} {392, 332}

\bibitem[\protect\citeauthoryear{{Loose}, {Kruegel}  \& {Tutukov}}{{Loose}
  et~al.}{1982}]{Loose1982}
{Loose} H.~H.,  {Kruegel} E.,   {Tutukov} A.,  1982, \aap, \href
  {http://adsabs.harvard.edu/abs/1982A%26A...105..342L} {105, 342}

\bibitem[\protect\citeauthoryear{{Lu}, {Do}, {Ghez}, {Morris}, {Yelda}  \&
  {Matthews}}{{Lu} et~al.}{2013}]{Lu2013}
{Lu} J.~R.,  {Do} T.,  {Ghez} A.~M.,  {Morris} M.~R.,  {Yelda} S.,   {Matthews}
  K.,  2013, \mn@doi [\apj] {10.1088/0004-637X/764/2/155}, \href
  {http://adsabs.harvard.edu/abs/2013ApJ...764..155L} {764, 155}

\bibitem[\protect\citeauthoryear{{MacLeod}, {Ramirez-Ruiz}, {Grady}  \&
  {Guillochon}}{{MacLeod} et~al.}{2013}]{MacLeod2013}
{MacLeod} M.,  {Ramirez-Ruiz} E.,  {Grady} S.,   {Guillochon} J.,  2013,
  \mn@doi [\apj] {10.1088/0004-637X/777/2/133}, \href
  {https://ui.adsabs.harvard.edu/abs/2013ApJ...777..133M} {777, 133}

\bibitem[\protect\citeauthoryear{{Madigan}, {Halle}, {Moody}, {McCourt},
  {Nixon}  \& {Wernke}}{{Madigan} et~al.}{2018}]{Madigan2018}
{Madigan} A.-M.,  {Halle} A.,  {Moody} M.,  {McCourt} M.,  {Nixon} C.,
  {Wernke} H.,  2018, \mn@doi [\apj] {10.3847/1538-4357/aaa714}, \href
  {https://ui.adsabs.harvard.edu/abs/2018ApJ...853..141M} {853, 141}

\bibitem[\protect\citeauthoryear{{Magorrian} \& {Tremaine}}{{Magorrian} \&
  {Tremaine}}{1999}]{Magorrian1999}
{Magorrian} J.,  {Tremaine} S.,  1999, \mn@doi [\mnras]
  {10.1046/j.1365-8711.1999.02853.x}, \href
  {https://ui.adsabs.harvard.edu/abs/1999MNRAS.309..447M} {309, 447}

\bibitem[\protect\citeauthoryear{Merloni et~al.,}{Merloni
  et~al.}{2012}]{Merloni2012}
Merloni A.,  et~al., 2012, eROSITA Science Book: Mapping the Structure of the
  Energetic Universe (\mn@eprint {arXiv} {arXiv:1209.3114v2})

\bibitem[\protect\citeauthoryear{{Merritt}}{{Merritt}}{2013}]{Merritt2013}
{Merritt} D.,  2013, {Dynamics and Evolution of Galactic Nuclei}

\bibitem[\protect\citeauthoryear{{Mockler}, {Twum}, {Auchettl}, {Dodd},
  {French}, {Law-Smith}  \& {Ramirez-Ruiz}}{{Mockler}
  et~al.}{2021}]{Mockler2021}
{Mockler} B.,  {Twum} A.~A.,  {Auchettl} K.,  {Dodd} S.,  {French} K.~D.,
  {Law-Smith} J. A.~P.,   {Ramirez-Ruiz} E.,  2021, arXiv e-prints, \href
  {https://ui.adsabs.harvard.edu/abs/2021arXiv211003013M} {p. arXiv:2110.03013}

\bibitem[\protect\citeauthoryear{{Morris}}{{Morris}}{1993}]{Morris1993}
{Morris} M.,  1993, \mn@doi [\apj] {10.1086/172607}, \href
  {http://adsabs.harvard.edu/abs/1993ApJ...408..496M} {408, 496}

\bibitem[\protect\citeauthoryear{{Neumayer}, {Seth}  \& {B{\"o}ker}}{{Neumayer}
  et~al.}{2020}]{Neumayer2020}
{Neumayer} N.,  {Seth} A.,   {B{\"o}ker} T.,  2020, \mn@doi [\aapr]
  {10.1007/s00159-020-00125-0}, \href
  {https://ui.adsabs.harvard.edu/abs/2020A&ARv..28....4N} {28, 4}

\bibitem[\protect\citeauthoryear{{Pestoni}, {Bortolas}, {Capelo}  \&
  {Mayer}}{{Pestoni} et~al.}{2021}]{Pestoni2021}
{Pestoni} B.,  {Bortolas} E.,  {Capelo} P.~R.,   {Mayer} L.,  2021, \mn@doi
  [\mnras] {10.1093/mnras/staa3496}, \href
  {https://ui.adsabs.harvard.edu/abs/2021MNRAS.500.4628P} {500, 4628}

\bibitem[\protect\citeauthoryear{{Pfister}, {Volonteri}, {Dai}  \&
  {Colpi}}{{Pfister} et~al.}{2020}]{Pfister2020}
{Pfister} H.,  {Volonteri} M.,  {Dai} J.~L.,   {Colpi} M.,  2020, \mn@doi
  [\mnras] {10.1093/mnras/staa1962}, \href
  {https://ui.adsabs.harvard.edu/abs/2020MNRAS.tmp.2074P} {}

\bibitem[\protect\citeauthoryear{{Pfister}, {Toscani}, {Wong}, {Dai}, {Lodato}
  \& {Rossi}}{{Pfister} et~al.}{2022}]{Pfister2022}
{Pfister} H.,  {Toscani} M.,  {Wong} T. H.~T.,  {Dai} J.~L.,  {Lodato} G.,
  {Rossi} E.~M.,  2022, \mn@doi [\mnras] {10.1093/mnras/stab3387}, \href
  {https://ui.adsabs.harvard.edu/abs/2022MNRAS.510.2025P} {510, 2025}

\bibitem[\protect\citeauthoryear{{Preto} \& {Amaro-Seoane}}{{Preto} \&
  {Amaro-Seoane}}{2010}]{Preto2010}
{Preto} M.,  {Amaro-Seoane} P.,  2010, \mn@doi [\apjl]
  {10.1088/2041-8205/708/1/L42}, \href
  {http://adsabs.harvard.edu/abs/2010ApJ...708L..42P} {708, L42}

\bibitem[\protect\citeauthoryear{{Rau} et~al.,}{{Rau} et~al.}{2009}]{Rau2009}
{Rau} A.,  et~al., 2009, \mn@doi [\pasp] {10.1086/605911}, \href
  {https://ui.adsabs.harvard.edu/abs/2009PASP..121.1334R} {121, 1334}

\bibitem[\protect\citeauthoryear{{Rees}}{{Rees}}{1988}]{Rees1988}
{Rees} M.~J.,  1988, \mn@doi [\nat] {10.1038/333523a0}, \href
  {https://ui.adsabs.harvard.edu/abs/1988Natur.333..523R} {333, 523}

\bibitem[\protect\citeauthoryear{{Rossi}, {Stone}, {Law-Smith}, {MacLeod},
  {Lodato}, {Dai}  \& {Mand el}}{{Rossi} et~al.}{2020}]{Rossi2020}
{Rossi} E.~M.,  {Stone} N.~C.,  {Law-Smith} J. A.~P.,  {MacLeod} M.,  {Lodato}
  G.,  {Dai} J.~L.,   {Mand el} I.,  2020, arXiv e-prints, \href
  {https://ui.adsabs.harvard.edu/abs/2020arXiv200512528R} {p. arXiv:2005.12528}

\bibitem[\protect\citeauthoryear{{Ryu}, {Krolik}, {Piran}  \& {Noble}}{{Ryu}
  et~al.}{2020}]{Ryu2020eta}
{Ryu} T.,  {Krolik} J.,  {Piran} T.,   {Noble} S.~C.,  2020, \mn@doi [\apj]
  {10.3847/1538-4357/abb3cf}, \href
  {https://ui.adsabs.harvard.edu/abs/2020ApJ...904...98R} {904, 98}

\bibitem[\protect\citeauthoryear{{Sanders}}{{Sanders}}{1998}]{Sanders1998}
{Sanders} R.~H.,  1998, \mn@doi [\mnras] {10.1046/j.1365-8711.1998.01127.x},
  \href {http://adsabs.harvard.edu/abs/1998MNRAS.294...35S} {294, 35}

\bibitem[\protect\citeauthoryear{{Sch{\"o}del}, {Gallego-Cano}, {Dong},
  {Nogueras-Lara}, {Gallego-Calvente}, {Amaro-Seoane}  \&
  {Baumgardt}}{{Sch{\"o}del} et~al.}{2017}]{Schodel2017}
{Sch{\"o}del} R.,  {Gallego-Cano} E.,  {Dong} H.,  {Nogueras-Lara} F.,
  {Gallego-Calvente} A.~T.,  {Amaro-Seoane} P.,   {Baumgardt} H.,  2017,
  preprint, \href {http://adsabs.harvard.edu/abs/2017arXiv170103817S} {}
  (\mn@eprint {arXiv} {1701.03817})

\bibitem[\protect\citeauthoryear{{Sersic}}{{Sersic}}{1968}]{Sersic1968}
{Sersic} J.~L.,  1968, {Atlas de Galaxias Australes}

\bibitem[\protect\citeauthoryear{{Shappee} et~al.,}{{Shappee}
  et~al.}{2014}]{Shappee2014}
{Shappee} B.~J.,  et~al., 2014, \mn@doi [\apj] {10.1088/0004-637X/788/1/48},
  \href {https://ui.adsabs.harvard.edu/abs/2014ApJ...788...48S} {788, 48}

\bibitem[\protect\citeauthoryear{{Spera}, {Mapelli}  \& {Bressan}}{{Spera}
  et~al.}{2015}]{Spera2015}
{Spera} M.,  {Mapelli} M.,   {Bressan} A.,  2015, \mn@doi [\mnras]
  {10.1093/mnras/stv1161}, \href
  {http://adsabs.harvard.edu/abs/2015MNRAS.451.4086S} {451, 4086}

\bibitem[\protect\citeauthoryear{{Spitzer} \& {Hart}}{{Spitzer} \&
  {Hart}}{1971}]{Spitzer1971}
{Spitzer} Lyman J.,  {Hart} M.~H.,  1971, \mn@doi [\apj] {10.1086/150855},
  \href {https://ui.adsabs.harvard.edu/abs/1971ApJ...164..399S} {164, 399}

\bibitem[\protect\citeauthoryear{{Stegmann}, {Capelo}, {Bortolas}  \&
  {Mayer}}{{Stegmann} et~al.}{2020}]{Stegmann2020}
{Stegmann} J.,  {Capelo} P.~R.,  {Bortolas} E.,   {Mayer} L.,  2020, \mn@doi
  [\mnras] {10.1093/mnras/staa170}, \href
  {https://ui.adsabs.harvard.edu/abs/2020MNRAS.492.5247S} {492, 5247}

\bibitem[\protect\citeauthoryear{{Stone} \& {Metzger}}{{Stone} \&
  {Metzger}}{2016}]{Stone2016}
{Stone} N.~C.,  {Metzger} B.~D.,  2016, \mn@doi [\mnras]
  {10.1093/mnras/stv2281}, \href
  {https://ui.adsabs.harvard.edu/abs/2016MNRAS.455..859S} {455, 859}

\bibitem[\protect\citeauthoryear{{Stone} \& {van Velzen}}{{Stone} \& {van
  Velzen}}{2016}]{Stone2016obs}
{Stone} N.~C.,  {van Velzen} S.,  2016, \mn@doi [\apjl]
  {10.3847/2041-8205/825/1/L14}, \href
  {https://ui.adsabs.harvard.edu/abs/2016ApJ...825L..14S} {825, L14}

\bibitem[\protect\citeauthoryear{{Stone}, {Generozov}, {Vasiliev}  \&
  {Metzger}}{{Stone} et~al.}{2018}]{Stone2018}
{Stone} N.~C.,  {Generozov} A.,  {Vasiliev} E.,   {Metzger} B.~D.,  2018,
  \mn@doi [\mnras] {10.1093/mnras/sty2045}, \href
  {https://ui.adsabs.harvard.edu/abs/2018MNRAS.480.5060S} {480, 5060}

\bibitem[\protect\citeauthoryear{{Stone}, {Vasiliev}, {Kesden}, {Rossi},
  {Perets}  \& {Amaro-Seoane}}{{Stone} et~al.}{2020}]{Stone2020}
{Stone} N.~C.,  {Vasiliev} E.,  {Kesden} M.,  {Rossi} E.~M.,  {Perets} H.~B.,
  {Amaro-Seoane} P.,  2020, \mn@doi [\ssr] {10.1007/s11214-020-00651-4}, \href
  {https://ui.adsabs.harvard.edu/abs/2020SSRv..216...35S} {216, 35}

\bibitem[\protect\citeauthoryear{{Tang}, {Bressan}, {Rosenfield}, {Slemer},
  {Marigo}, {Girardi}  \& {Bianchi}}{{Tang} et~al.}{2014}]{Tang2014}
{Tang} J.,  {Bressan} A.,  {Rosenfield} P.,  {Slemer} A.,  {Marigo} P.,
  {Girardi} L.,   {Bianchi} L.,  2014, \mn@doi [\mnras]
  {10.1093/mnras/stu2029}, \href
  {http://adsabs.harvard.edu/abs/2014MNRAS.445.4287T} {445, 4287}

\bibitem[\protect\citeauthoryear{{Tout}, {Pols}, {Eggleton}  \& {Han}}{{Tout}
  et~al.}{1996}]{Tout1996}
{Tout} C.~A.,  {Pols} O.~R.,  {Eggleton} P.~P.,   {Han} Z.,  1996, \mn@doi
  [\mnras] {10.1093/mnras/281.1.257}, \href
  {https://ui.adsabs.harvard.edu/abs/1996MNRAS.281..257T} {281, 257}

\bibitem[\protect\citeauthoryear{{Toyouchi}, {Inayoshi}, {Ishigaki}  \&
  {Tominaga}}{{Toyouchi} et~al.}{2021}]{Toyouchi2021}
{Toyouchi} D.,  {Inayoshi} K.,  {Ishigaki} M.~N.,   {Tominaga} N.,  2021, arXiv
  e-prints, \href {https://ui.adsabs.harvard.edu/abs/2021arXiv211206151T} {p.
  arXiv:2112.06151}

\bibitem[\protect\citeauthoryear{{Tremaine}, {Ostriker}  \&
  {Spitzer}}{{Tremaine} et~al.}{1975}]{Tremaine1975}
{Tremaine} S.~D.,  {Ostriker} J.~P.,   {Spitzer} Jr. L.,  1975, \mn@doi [\apj]
  {10.1086/153422}, \href {http://adsabs.harvard.edu/abs/1975ApJ...196..407T}
  {196, 407}

\bibitem[\protect\citeauthoryear{{Vasiliev}}{{Vasiliev}}{2017}]{Vasiliev2017}
{Vasiliev} E.,  2017, \mn@doi [\apj] {10.3847/1538-4357/aa8cc8}, \href
  {http://adsabs.harvard.edu/abs/2017ApJ...848...10V} {848, 10}

\bibitem[\protect\citeauthoryear{{Vasiliev}}{{Vasiliev}}{2019}]{Vasiliev2019}
{Vasiliev} E.,  2019, \mn@doi [\mnras] {10.1093/mnras/sty2672}, \href
  {https://ui.adsabs.harvard.edu/abs/2019MNRAS.482.1525V} {482, 1525}

\bibitem[\protect\citeauthoryear{{V{\'a}zquez-Aceves}, {Zwick}, {Bortolas},
  {Capelo}, {Seoane}, {Mayer}  \& {Chen}}{{V{\'a}zquez-Aceves}
  et~al.}{2021}]{Vazquez-Aceves2021}
{V{\'a}zquez-Aceves} V.,  {Zwick} L.,  {Bortolas} E.,  {Capelo} P.~R.,
  {Seoane} P.~A.,  {Mayer} L.,   {Chen} X.,  2021, \mn@doi [\mnras]
  {10.1093/mnras/stab3485}, \href
  {https://ui.adsabs.harvard.edu/abs/2021MNRAS.tmp.3163V} {}

\bibitem[\protect\citeauthoryear{{Wang} \& {Merritt}}{{Wang} \&
  {Merritt}}{2004}]{Wang2004}
{Wang} J.,  {Merritt} D.,  2004, \mn@doi [\apj] {10.1086/379767}, \href
  {https://ui.adsabs.harvard.edu/abs/2004ApJ...600..149W} {600, 149}

\bibitem[\protect\citeauthoryear{{Zhang}, {Romano}, {Ivison}, {Papadopoulos}
  \& {Matteucci}}{{Zhang} et~al.}{2018}]{Zhang2018}
{Zhang} Z.-Y.,  {Romano} D.,  {Ivison} R.~J.,  {Papadopoulos} P.~P.,
  {Matteucci} F.,  2018, \mn@doi [\nat] {10.1038/s41586-018-0196-x}, \href
  {http://adsabs.harvard.edu/abs/2018Natur.558..260Z} {558, 260}

\bibitem[\protect\citeauthoryear{{Zhu}, {Vasiliev}, {Li}  \& {Jing}}{{Zhu}
  et~al.}{2018}]{Zhu2018}
{Zhu} Q.,  {Vasiliev} E.,  {Li} Y.,   {Jing} Y.,  2018, \mn@doi [\mnras]
  {10.1093/mnras/sty079}, \href
  {https://ui.adsabs.harvard.edu/abs/2018MNRAS.476....2Z} {476, 2}

\bibitem[\protect\citeauthoryear{{Zwick}, {Capelo}, {Bortolas}, {Mayer}  \&
  {Amaro-Seoane}}{{Zwick} et~al.}{2020}]{Zwick2020}
{Zwick} L.,  {Capelo} P.~R.,  {Bortolas} E.,  {Mayer} L.,   {Amaro-Seoane} P.,
  2020, \mn@doi [\mnras] {10.1093/mnras/staa1314}, \href
  {https://ui.adsabs.harvard.edu/abs/2020MNRAS.495.2321Z} {495, 2321}

\bibitem[\protect\citeauthoryear{{Zwick}, {Capelo}, {Bortolas},
  {V{\'a}zquez-Aceves}, {Mayer}  \& {Amaro-Seoane}}{{Zwick}
  et~al.}{2021}]{Zwick2021}
{Zwick} L.,  {Capelo} P.~R.,  {Bortolas} E.,  {V{\'a}zquez-Aceves} V.,  {Mayer}
  L.,   {Amaro-Seoane} P.,  2021, \mn@doi [\mnras] {10.1093/mnras/stab1818},
  \href {https://ui.adsabs.harvard.edu/abs/2021MNRAS.506.1007Z} {506, 1007}

\makeatother
\end{thebibliography}


\bsp	
\label{lastpage}
\end{document}